\documentclass[nofootinbib,showkeys,twocolumn,a4paper,10pt,aps,prd,preprintnumbers,superscriptaddress,floatfix]{revtex4-2}
\pdfoutput=1
\usepackage{amsmath,amssymb,graphicx,bm,slashed,array,multirow}
\usepackage[dvipsnames]{xcolor}
\usepackage{tikz}
\usepackage{lineno,hyperref}
\hypersetup{
    colorlinks=true,
    linkcolor=blue,
    citecolor=blue,
    filecolor=magenta,
    urlcolor=cyan,
    pdftitle={Projected sensitivity of the ANUBIS detector to photophobic and fermion-coupled axion-like particles},
    pdfpagemode=UseOutlines
}
\usepackage{xspace}
\usepackage{siunitx}
\usepackage{enumitem}
\setlist{itemsep=0mm}
\newcommand{\eg}{\mbox{\itshape e.g.}\xspace}
\newcommand{\ie}{\mbox{\itshape i.e.}\xspace}

\newcommand{\anubis}{\mbox{ANUBIS}\xspace}
\newcommand{\setanubis}{\mbox{\textsc{SET-ANUBIS}}\xspace}

\newcommand{\atlas}{\mbox{ATLAS}\xspace}

\newcommand{\mathusla}{\mbox{MATHUSLA}\xspace}

\newcommand{\pythia}[1][]{\mbox{\textsc{Pythia#1}}\xspace}

\newcommand{\madspin}{\mbox{\textsc{MadSpin}}\xspace}
\newcommand{\madversion}[1]{MG5\_aMC@NLO #1}

\newcommand{\metre}{\ensuremath{\textnormal{m}}\xspace}

\newcommand{\GeV}{\ensuremath{\textnormal{GeV}}\xspace}
\newcommand{\TeV}{\ensuremath{\textnormal{TeV}}\xspace}
\newcommand{\pb}{\ensuremath{\rm pb}\xspace}

\newcommand{\invfb}{\ensuremath{{\rm fb}^{-1}}\xspace}

\newcommand{\invab}{\ensuremath{{\rm ab}^{-1}}\xspace}

\newcommand{\met}{\ensuremath{E_T^\text{miss}}\xspace} % Missing Transverse Energy
\newcommand{\order}[1]{\ensuremath{\mathcal{O}(#1)}\xspace} % Order of X
\let\BR=\br
\newcommand{\Ca}{\ensuremath{C_{a\Phi}}\xspace}
\newcommand{\CZh}{\ensuremath{C_{Zh}^{\mathrm{eff}}}\xspace}
\newcommand{\CZseven}{\ensuremath{C_{Zh}^{(7)}}\xspace}
\newcommand{\Lag}{\ensuremath{\mathcal{L}}\xspace}

\def\pmbanner{}
\begin{document}
\title{\pmbanner{\LARGE \texorpdfstring{Projected sensitivity of the \anubis detector to photophobic and fermion-coupled axion-like particles\\
\includegraphics[width=0.1\linewidth]{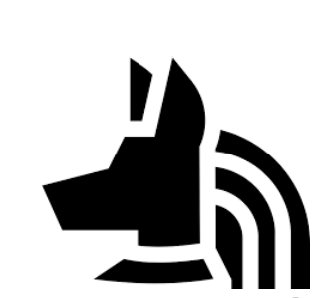}\\
{\large\textbf{ANUBIS Collaboration}}
\\[-0.2em]
{\normalsize\textit{E-mail: }\href{mailto:anubis-publications@cern.ch}{anubis-publications@cern.ch}}
\\[-0.2em]
}{Projected sensitivity of the ANUBIS detector to photophobic and fermion-coupled axion-like particles}}}

\begin{abstract}
\textit{Abstract:}
Axion-like particles (ALPs) are a common benchmark for experiments at the lifetime frontier, as their suppressed couplings to Standard Model particles can render them long-lived on detector scales.
This article presents the projected sensitivity of the \anubis detector at the High-Luminosity LHC (HL-LHC) to two scenarios: a photophobic ALP model and a modified version of the flavour-diagonal, fermion-coupled ALP benchmark of the Physics Beyond Colliders initiative that introduces an additional $h\to Za$ decay.
%In both models the ALP is produced in association with an electroweak vector boson, $W^\pm$ or $Z$, which provides a prompt signal that can be associated with the long-lived ALP and allows for an effectively background-free search with \anubis.
%The complete workflow --- from event generation through lifetime reweighting, detector acceptance and event selection to contour construction --- is performed with v1.0.0 of the \setanubis framework.
% Direct associated production is too strongly suppressed by the mass-proportional ALP coupling to light initial-state quarks to yield an exclusion reach.
% For a strictly fermion-coupled ALP, the top-loop-induced Higgs channel reaches a maximum of $3.63\pm0.39$ expected events and remains just below the conservative four-event threshold adopted for a background-free 95\% confidence-level projection.
%
For the photophobic case, the \anubis detector will reach a peak sensitivity of ${\sigma_{Wa}\cdot\BR(a\rightarrow gg)=6.2\times10^{-4}~\pb}$ at $c\tau=1.95~\metre$ for an ALP mass of $10~\GeV$.
This will improve on the current exclusion limits from the \atlas experiment by two orders of magnitude and extend the $c\tau$ coverage by three orders of magnitude.
For the fermion-coupled case with a dimension-seven Higgs--ALP interaction, the peak sensitivity to the fermion coupling is $\Ca=4.4\times10^{-8}$, reached for ALP masses between $16$ and $23~\GeV$.
Overall, the \anubis detector is shown to provide unique coverage of the parameter space of both models, in regions that can be probed neither by the main LHC experiments nor by forward detectors at the HL-LHC.
   \begin{center}
       %Version 2.0\\[-0.3em]
       \today
   \end{center}
   \vspace{-0.5cm}
\end{abstract}

% authors (by surname)
\author{Riley~Aum\^o{}nier}
\affiliation{Formerly at: Cavendish Laboratory, University of Cambridge, Cambridge, United Kingdom}
\author{Oleg~Brandt}
\affiliation{Cavendish Laboratory, University of Cambridge, Cambridge, United Kingdom}
\author{Th\'e{}o~Reymermier}
\affiliation{Formerly at: Cavendish Laboratory, University of Cambridge, Cambridge, United Kingdom}
\affiliation{Université Lyon 1, CNRS, IP2I, UMR 5822, Villeurbanne, France}
\author{Francis~Sayer}
\affiliation{Formerly at: Cavendish Laboratory, University of Cambridge, Cambridge, United Kingdom}
\author{and Paul~Swallow}
\affiliation{Cavendish Laboratory, University of Cambridge, Cambridge, United Kingdom}
\author{for the ANUBIS Collaboration}
\noaffiliation

\def\andname{\hspace*{-0.5em}}
\maketitle

%%%%%%%%%%%%%%%%%%%%%%%%%%%%%%%%%%%%%%%
\section{Introduction}
\label{sec:intro}
%%%%%%%%%%%%%%%%%%%%%%%%%%%%%%%%%%%%%%%
Long-lived particles (LLPs) provide a direct experimental handle on feebly coupled extensions of the Standard Model (SM).
Their macroscopic decay lengths can arise from small portal couplings, approximate symmetries, or interactions suppressed by a high scale.
The resulting signatures are difficult to capture with detectors and reconstruction algorithms designed primarily for prompt decays. This has motivated a range of dedicated LLP experiments at the LHC lifetime frontier~\cite{Antel:2023hkf}.

\begin{figure}[ht]
    \centering
    \begin{tikzpicture}
        \node[inner sep=0] (image) at (0,0) {\includegraphics[width=0.75\linewidth]{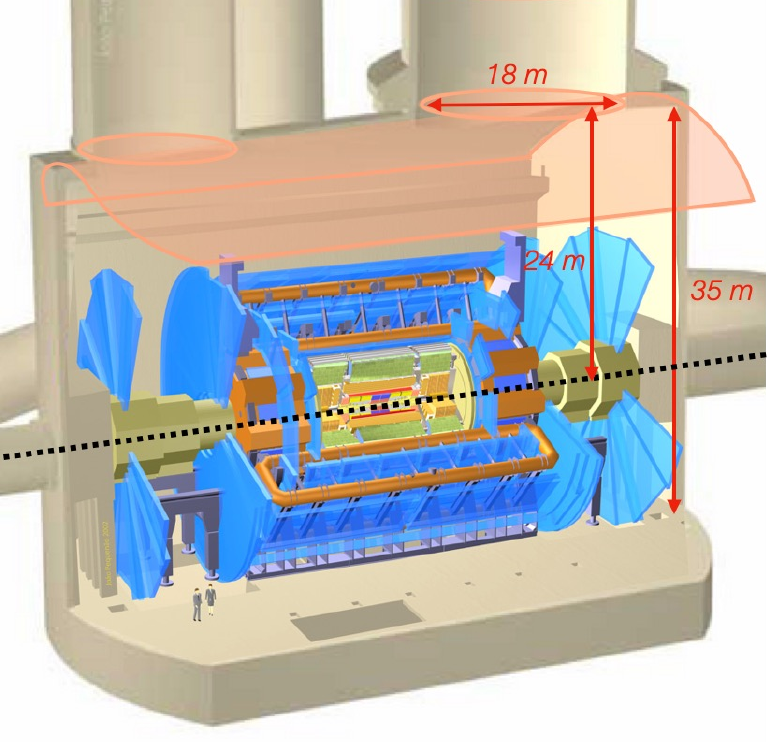}};
        \node[inner sep=0] at (1.1,2.9) {{\color{red}{\textbf{PX14}}}};
        \node[inner sep=0] at (-2,2.9) {{\color{red}{\textbf{PX16}}}};
     \end{tikzpicture}
    \caption{
    Layout of the \atlas underground cavern UX15 at LHC Point 1, including the access shafts PX14 and PX16. 
    The area highlighted in orange illustrates the tracking layers of the \anubis detector.}
    \label{fig:SketchATLAS_undergroundCavern}
\end{figure}

The \anubis detector is proposed to instrument the ceiling of the \atlas UX15 cavern with tracking layers of Resistive Plate Chambers (RPCs)~\cite{Bauer:2019vqk,ANUBIS:2025sgg}, as shown in Figure~\ref{fig:SketchATLAS_undergroundCavern}.
It would use the air-filled region above \atlas as a decay volume and reconstruct electrically neutral long-lived particles through their charged decay products.
The transverse placement of the detector gives access to particles produced at electroweak partonic centre-of-mass energies, while integration with the \atlas trigger and data-acquisition systems allows the displaced decay to be studied together with the associated prompt component of the event.
The detector layout, fiducial volume, expected backgrounds, and baseline reconstruction strategy are described in Ref.~\cite{ANUBIS:2025sgg}.

Axion-like particles (ALPs) are pseudo-Nambu--Goldstone bosons associated with the spontaneous breaking of an approximate global symmetry~\cite{Choi:2020rgn}.
Their approximate shift symmetry naturally leads to derivative interactions suppressed by a scale, $\Lambda$, and a sufficiently small coupling can give an ALP a macroscopic lifetime.
The Physics Beyond Colliders (PBC) initiative includes three ALP benchmarks: BC9, where the ALP couples to photons; BC10, where the ALP couples to fermions; and BC11, where the ALP couples to gluons~\cite{Beacham:2019nyx,Alemany:2019vsk}.
In the course of this study, BC10 and BC11 were evaluated for an ALP produced in association with a $W^\pm$ or $Z$ boson, and \anubis was found to have no sensitivity to either benchmark.
This article therefore considers instead a gluon-coupled `photophobic' scenario with suppressed photon couplings~\cite{Craig:2018kne,Brivio:2017ije}, a benchmark widely adopted by the \atlas Collaboration~\cite{ATLAS:2025pak}, and a fermion-coupled BC10 model extended by an exotic Higgs decay, $h\to Za$~\cite{Bauer:2017ris,Curtin:2018mvb}, referred to as `BC10h'.
In both cases the ALP is produced in association with a $W^\pm$ or $Z$ boson, which, through their leptonic decays, provide a natural trigger for \atlas and allow the backgrounds to be reduced to negligible levels.

% It is also expected to be strongly suppressed in a universal fermion-coupling benchmark, since proton parton luminosities are dominated by light quarks whereas the physical ALP coupling is proportional to the fermion mass.
% The decay $h\to Za$ avoids the light-quark suppression through resonant Higgs production.
% Within the strict dimension-five fermion benchmark it is generated mainly by a top-quark loop.
% A second interpretation is therefore considered in which the ALP lifetime remains controlled by its fermion coupling, while production receives an independent contribution from the leading polynomial Higgs--ALP operator, which occurs at dimension seven in the linearly realised electroweak effective theory~\cite{Bauer:2017ris}.

Previous \anubis sensitivity studies covered exotic decays of the Higgs boson to long-lived scalars~\cite{ANUBIS:2025sgg} and heavy neutral leptons coupled to electrons or muons~\cite{ANUBIS:HNLemu2026}, the latter performed with a pre-release version of the \setanubis framework, a modular, configuration-driven pipeline for \anubis sensitivity studies~\cite{erner2026setanubismodularpipelineanubis}.
% [OB] Add the BC8 (tau-coupled HNL) paper here once it is public, e.g. "... and to taus~\cite{...}".
All stages of the present analysis are performed with the released \setanubis~v1.0.0 software~\cite{Reymermier:SETANUBISv100}.
A common workflow is particularly important here, because the production rate, decay branching fractions, lifetime, geometry, and selection efficiencies vary over several orders of magnitude across the two-dimensional parameter scans; it also ensures that the two models are treated consistently.

Existing constraints on the two scenarios are of different nature.
For photophobic ALPs, the most directly comparable result is the \atlas search for a single displaced vertex in the muon spectrometer, which was interpreted for a long-lived photophobic ALP produced in association with a $W^\pm$ or $Z$ boson and decaying to gluons, for ALP masses, $m_a$, between 0.1 and $40~\GeV$~\cite{ATLAS:2025pak}.
Other \atlas searches, for displaced jets in the calorimeter~\cite{bib:calratio} and for displaced vertices in the inner detector~\cite{bib:hdmi}, target shorter $c\tau$ for similar $m_a$.
For heavier, promptly decaying photophobic ALPs, constraints have been derived by reinterpreting LEP and LHC searches~\cite{Craig:2018kne,Aiko:2024xiv}.
In the BC10h scenario, the exotic decay $h\to Za$ is constrained directly by \atlas searches for a light, hadronically decaying resonance produced in association with a $Z$ boson~\cite{ATLAS:2020hZaHad,ATLAS:2024hZaHad} and by searches for Higgs boson decays to a $Z$ boson and a new light boson in four-lepton final states~\cite{ATLAS:2021H4l}, and indirectly by global fits of the Higgs boson couplings that bound its branching fraction into undetected final states~\cite{ATLAS:2016neq}.
Fermion-coupled ALPs below a few \GeV are, in addition, constrained by flavour and beam-dump experiments~\cite{Beacham:2019nyx,Garcia:2023awt}.
All of these searches target promptly decaying or short-lived ALPs, or masses below the region of interest here, and are therefore complementary to the long-lived regime probed by \anubis.

Section~\ref{sec:theory} defines the effective ALP Lagrangian and the two ALP models considered.
Section~\ref{sec:analysis} describes the event generation and \setanubis analysis.
The projected yields and sensitivity contours are presented in Section~\ref{sec:results}, and conclusions are drawn in Section~\ref{sec:conclusion}.

%%%%%%%%%%%%%%%%%%%%%%%%%%%%%%%%%%%%%%%
\section{Axion-like particles}
\label{sec:theory}
%%%%%%%%%%%%%%%%%%%%%%%%%%%%%%%%%%%%%%%
\subsection{Effective Lagrangian}
\label{sec:eft}
An ALP is described at energies below the symmetry-breaking scale by an Effective Field Theory (EFT) organised in inverse powers of an energy scale, $\Lambda$.
For a linearly realised electroweak symmetry, the shift-symmetric interactions containing one ALP field up to mass dimension five may be written as~\cite{Brivio:2017ije}
\begin{align}
\Lag_{\mathrm{eff}} ={}& \Lag_{\mathrm{SM}}
+\frac{1}{2}(\partial_\mu a)(\partial^\mu a)
-\frac{1}{2}m_a^2a^2 \nonumber\\
&+i\Ca\frac{\partial^\mu a}{\Lambda}
\left(\Phi^\dagger\overleftrightarrow{D}_\mu\Phi\right) \nonumber\\
&-C_{\tilde{G}}\frac{a}{\Lambda}
 G^A_{\mu\nu}\widetilde G^{A\mu\nu} \nonumber\\
&-C_{\tilde{W}}\frac{a}{\Lambda}
 W^I_{\mu\nu}\widetilde W^{I\mu\nu} \nonumber\\
&-C_{\tilde{B}}\frac{a}{\Lambda}
 B_{\mu\nu}\widetilde B^{\mu\nu}
+\ldots,
\label{eq:general_eft}
\end{align}
where
\begin{equation}
\Phi^\dagger\overleftrightarrow{D}_\mu\Phi
\equiv
\Phi^\dagger D_\mu\Phi
-
(D_\mu\Phi)^\dagger\Phi,
\end{equation}
and $\widetilde X^{\mu\nu} =\epsilon^{\mu\nu\rho\sigma}X_{\rho\sigma}/2$. The coefficients $C_{\tilde{G}}$, $C_{\tilde{W}}$, and $C_{\tilde{B}}$ parametrise the ALP couplings to the gluon and electroweak gauge fields, while \Ca is the ALP coupling to fermions and multiplies the Higgs-current operator~\cite{Brivio:2017ije}. The mass term softly breaks the ALP shift symmetry, and the ALP mass, $m_a$, is treated as an independent parameter.

After Electroweak Symmetry Breaking (EWSB), the interaction with fermions can be expressed as
%and rotation to the fermion mass basis, the flavour-diagonal vector currents are conserved, and the physical derivative interaction with fermions can be expressed through axial currents,
\begin{equation}
%\Lag_{aff}=\sum_f \frac{c_{ff}}{2}\frac{\partial_\mu a}{\Lambda}
%\overline f\gamma^\mu\gamma_5 f,
\Lag_{aff}=-\frac{1}{2}\frac{\partial_\mu a}{\Lambda}\Ca\sum_{\psi=Q,L}
\overline \psi\gamma^\mu\gamma_5 \sigma_3 \psi,
\label{eq:mass_basis}
\end{equation}
where $Q$ and $L$ denote the left-handed quark and lepton doublets, and $\sigma_3$ is the third Pauli matrix acting in weak-isospin space~\cite{Brivio:2017ije}. The coefficient \Ca is taken to be flavour universal. However, the $\sigma_3$ structure implies a relative sign between up-type fermions and down-type quarks or charged leptons. Rotating to the fermion mass eigenstates and writing the interaction in the convention
\begin{equation}
\Lag_{aff}
=
\sum_f
\frac{c_{ff}}{2}
\frac{\partial_\mu a}{\Lambda}
\overline f\gamma^\mu\gamma_5 f,
\end{equation}
this gives
\begin{equation}
c_{u_i u_i}=-\Ca,
\qquad
c_{d_i d_i}=c_{\ell_i\ell_i}=+\Ca,
\label{eq:cff_mapping}
\end{equation}
where $u_i=(u,c,t)$, $d_i=(d,s,b)$, and
$\ell_i=(e,\mu,\tau)$. In particular,
\begin{equation}
c_{tt}=-\Ca.
\label{eq:ctt_mapping}
\end{equation}
% where $f\in[u,d,s,c,t,b,e,\mu,\tau]$.
% Neutrino interactions do not appear at this order in the SM limit of vanishing neutrino masses~\cite{Bauer:2017ris}. For this study, the fermion couplings are assumed to be universal and flavour-diagonal, \ie $c_{ff}=\Ca$ for all charged fermions $f$.

Likewise, after EWSB, the interaction of the ALP with photons can be expressed as
\begin{equation}
\Lag_{a\gamma\gamma}=-\frac{1}{4}g_{a\gamma\gamma}aF_{\mu\nu}\tilde{F}^{\mu\nu},
\label{eq:ALPphotonInt}
\end{equation}
where $g_{a\gamma\gamma}$ is the coupling between the ALP and two photons,
\begin{equation}
g_{a\gamma\gamma}=\frac{4}{\Lambda}(C_{\tilde{B}}\cos^2\theta_W+C_{\tilde{W}}\sin^2\theta_W),
\label{eq:ALPphotonCoupling}
\end{equation}
and $\theta_W$ is the Weinberg angle~\cite{Brivio:2017ije}. Throughout this article, the energy scale $\Lambda=1~\TeV$ is fixed, reducing the free parameters to $m_a$ and the ALP couplings.

\subsection{Photophobic model}
\label{sec:photophobic}
Photon-coupled ALPs, for which $\Ca=C_{\tilde{G}}=0$ and $C_{\tilde{W}},C_{\tilde{B}}\neq0$, are already strongly constrained by collider, beam-dump, and astrophysical searches; see, \eg, Refs.~\cite{Bauer:2017ris,Bauer:2018uxu} and references therein.
\anubis is not optimised for detecting diphoton final states, since its tracking layers require charged particles.
Moreover, the absence of a signal of photon-coupled ALPs at the LHC so far~\cite{ATLAS:2020lbl,CMS:2018erd,CMS:2022zfd,ATLAS:2023etl,ATLAS:2023ian} motivates scenarios in which the photon coupling is suppressed, \ie photophobic ALPs~\cite{Craig:2018kne}.
An ALP is photophobic if $g_{a\gamma\gamma}=0$, which by Equation~\eqref{eq:ALPphotonCoupling} requires $C_{\tilde{B}}=-C_{\tilde{W}}\tan^2\theta_W$. 
The production cross-section for photophobic ALPs is proportional to $C_{\tilde{W}}^2$, so $C_{\tilde{W}}=1$ is chosen to maximise the simulated cross-section, consistent with the approach of the \atlas Collaboration~\cite{ATLAS:2025pak}. 
The universal coupling to fermions, \Ca, is set to zero, leaving $m_{a}$ and the coupling to gluons, $C_{\tilde{G}}$, as the free parameters.

\begin{figure}[ht]
\centering
\includegraphics[width=0.49\linewidth]{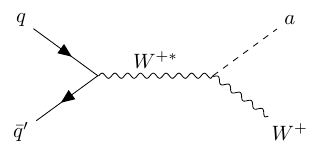}\hfill
\includegraphics[width=0.49\linewidth]{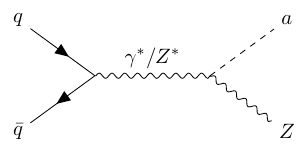}\\[1ex]
\caption{Feynman diagrams for the production of an ALP in association with a vector boson, $V=W^\pm,Z$, in the photophobic model.}
\label{fig:feyn}
\end{figure}

The ALP production modes in association with a vector boson in the photophobic scenario are shown in Figure~\ref{fig:feyn}.
The ALP subsequently decays into gluons, which hadronise into jets containing charged particles.
The signature is therefore a collimated set of charged-particle tracks from the resulting hadrons in the \anubis detector, accompanied by a prompt vector boson reconstructed in the \atlas detector.

For this model it is assumed that the ALP decays solely into a pair of gluons, with the partial width given by~\cite{ALPsEFTUFO,Brivio:2017ije}
\begin{equation}
    \Gamma(a\rightarrow gg)=\frac{2m_a^3}{\pi}\left|\frac{C_{\tilde{G}}}{\Lambda}\right|^2.
    \label{eq:GluonDecay}
\end{equation}

\subsection{BC10 model}
\label{sec:modBC10}

\begin{figure}[ht]
\centering
\includegraphics[width=0.40\linewidth]{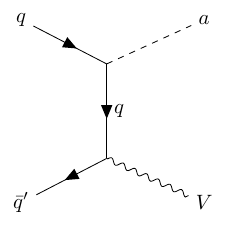}\hfill
\includegraphics[width=0.40\linewidth]{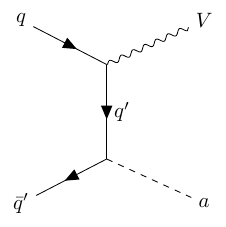}
\caption{Feynman diagrams for the production of an ALP in association with a vector boson, $V=W^\pm,Z$, in the BC10 model with fermionic ALP couplings, in addition to those in Figure~\ref{fig:feyn}.}
\label{fig:feynbc10}
\end{figure}

The BC10 benchmark is obtained by taking a universal, flavour-diagonal coefficient, \Ca,
and setting ${C_{\tilde{G}}=C_{\tilde{W}}=C_{\tilde{B}}=0}$ at the matching scale.
% The symbol \Ca is retained because it is the parameter used in the PBC benchmark implementation and in the sensitivity figures.
% Throughout this article $\Lambda=1~\TeV$ is fixed, so that the scan is specified by $(m_a,\Ca)$.
% Results for another value of $\Lambda$ can be translated at leading order through the ratio $\Ca/\Lambda$, provided the matching prescription and the running of the Wilson coefficients are kept fixed.
%
%Using the equations of motion,
The derivative interaction in Equation~\eqref{eq:mass_basis} can be written in the
chirality-flipping form~\cite{Brivio:2017ije},
\begin{equation}
\begin{aligned}
\Lag_{aff}
={}&
+i\Ca\frac{a}{\Lambda}
\sum_{u_i=u,c,t}
m_{u_i}\,\overline{u_i}\gamma_5u_i
\\
&-i\Ca\frac{a}{\Lambda}
\sum_{d_i=d,s,b}
m_{d_i}\,\overline{d_i}\gamma_5d_i
\\
&-i\Ca\frac{a}{\Lambda}
\sum_{\ell_i=e,\mu,\tau}
m_{\ell_i}\,\overline{\ell_i}\gamma_5\ell_i .
\end{aligned}
\label{eq:pseudoscalar}
\end{equation}
The relative sign between up-type fermions and down-type quarks or charged leptons follows directly from the $\sigma_3$ structure in Equation~\eqref{eq:mass_basis}. Equation~\eqref{eq:pseudoscalar} also makes explicit the fermion-mass scaling of the interaction, which strongly favours heavier fermions in production and decay.
The resulting Feynman diagrams for the production modes considered are shown in Figures~\ref{fig:feyn} and~\ref{fig:feynbc10}.
The free parameters of the model relevant for this study are $m_a$ and $\Ca$.
% Equation~\eqref{eq:pseudoscalar} is the origin of the strong preference for heavy fermions in both production and decay.
% For BC10, the explicit tree-level gauge coefficients are set to zero and loop-induced radiative ALP decays are not included in the signal normalisation.
% The loop-induced $hZa$ amplitude is retained separately, as discussed below.

\subsection{Extended BC10 model (BC10h)}
\label{sec:modBC10h}

% \subsection{Production channels}
% \label{sec:production}
Transverse detectors such as \anubis are particularly sensitive to LLPs produced in the decays of heavy states around the electroweak scale. The BC10 model from Section~\ref{sec:modBC10} is therefore extended to include resonant Higgs production, $pp\rightarrow h\rightarrow Za$, as was also considered for the transverse detector \mathusla~\cite{Curtin:2018mvb}.

\begin{figure}[ht]
\centering
\includegraphics[width=0.69\linewidth]{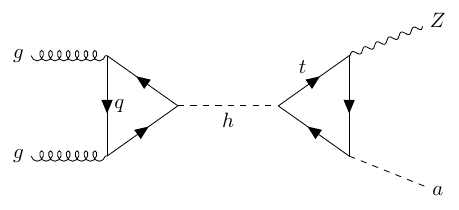}\\
\includegraphics[width=0.59\linewidth]{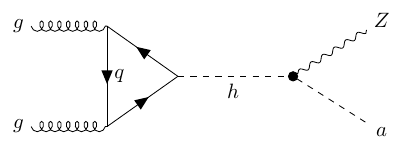}
\caption{
Feynman diagrams for the resonant production of an ALP in association with a $Z$ boson via a Higgs boson in the BC10h model, through (top) the top-quark loop and (bottom) the dimension-seven operator of Equation~\eqref{eq:d7}.
Only the dominant gluon-gluon fusion production mode of the Higgs boson is considered. 
}
\label{fig:feynres}
\end{figure}

No polynomial dimension-five operator generates ${h\to Za}$ at tree level in the linearly realised ALP EFT.
With only the interactions in Equation~\eqref{eq:mass_basis}, the leading amplitude is generated by a fermion loop and is dominated by the top quark~\cite{Bauer:2017ris}. 
The corresponding Feynman diagram is shown in Figure~\ref{fig:feynres}~(top).

The leading polynomial beyond-the-SM (BSM) interaction that generates ${h\to Za}$ at tree level in the ALP EFT occurs at dimension seven,
\begin{equation}
\Lag_{Zh}^{(7)}=
\frac{\CZseven}{\Lambda^3}(\partial^\mu a)
\left(\Phi^\dagger iD_\mu\Phi+\mathrm{h.c.}\right)
(\Phi^\dagger\Phi),
\label{eq:d7}
\end{equation}
where $\Phi$ is the SM Higgs doublet.
It is convenient to combine the loop-induced and higher-dimensional contributions into an effective coupling~\cite{Bauer:2017ris}. In the convention of Ref.~\cite{Bauer:2017ris}, the top-quark interaction is parametrised by $c_{tt}$, and the effective coupling is
\begin{align}
\CZh
&=
C_{Zh}^{(5)}
-\frac{N_c y_t^2}{8\pi^2}
T_3^t\,c_{tt}\,F
+\frac{v^2}{2\Lambda^2}\CZseven
\nonumber\\
&=
C_{Zh}^{(5)}
+\frac{N_c y_t^2}{8\pi^2}
T_3^t\,\Ca\,F
+\frac{v^2}{2\Lambda^2}\CZseven,
\label{eq:ceff}
\end{align}
where the second line follows from the mapping ${c_{tt}=-\Ca}$ of Equation~\eqref{eq:ctt_mapping}.
For the model considered here, $C_{Zh}^{(5)}=0$,
$N_c=3$, and $T_3^t=1/2$, and the Feynman diagram for the dimension-seven contribution is shown in Figure~\ref{fig:feynres}~(bottom). 
The loop function is
\begin{equation}
F=\int_0^1\!\mathrm{d}[xyz]\,
\frac{2m_t^2-xm_h^2-zm_Z^2}
{m_t^2-xy m_h^2-yz m_Z^2-xz m_a^2},
\label{eq:loopF}
\end{equation}
with $\mathrm{d}[xyz]=\mathrm{d}x\,\mathrm{d}y\,\mathrm{d}z\,\delta(1-x-y-z)$, and $F\simeq0.93$ over the mass range considered.
For $\Lambda=1~\TeV$, Equation~\eqref{eq:ceff} is numerically
\begin{equation}
\CZh\simeq+0.016\Ca+0.030\CZseven,
\label{eq:ceffnumeric}
\end{equation}
so the strict fermion-only limit is $\CZh\simeq+0.016\Ca$.

The partial decay width is given by
\begin{equation}
\Gamma(h\to Za)=\frac{m_h^3}{16\pi\Lambda^2}
|\CZh|^2\lambda^{3/2}\!\left(\frac{m_Z^2}{m_h^2},\frac{m_a^2}{m_h^2}\right),
\label{eq:hwidth}
\end{equation}
where $\lambda(x,y)=(1-x-y)^2-4xy$.
Since the $Z$ boson is on-shell, the decay is kinematically allowed only for ${m_a<m_h-m_Z}$.
The branching fraction and resonant production cross-section are then defined as
\begin{align}
\BR(h\to Za)&=\frac{\Gamma(h\to Za)}{\Gamma_h^{\mathrm{SM}}+\Gamma(h\to Za)},
\label{eq:hbr}\\
\sigma(pp\to h\to Za)&=\sigma(pp\to h)\,\BR(h\to Za),
\label{eq:hxsec}
\end{align}
where $\Gamma_h^{\mathrm{SM}}=4.088\times10^{-3}~\GeV$ for $m_h=125~\GeV$, and the gluon-gluon fusion production cross-section is ${\sigma(pp\to h)=54.67~\pb}$ at $\sqrt{s}=14~\TeV$~\cite{LHCHiggsCrossSectionWorkingGroup:2016ypw}. For BC10h, the free parameters are $m_a$, \CZh, and \Ca, where \CZh controls the production via the Higgs boson and \Ca controls the ALP decay width and hence the probability of the ALP reaching \anubis.

% Existing limits on universal fermion couplings are usually displayed in the $(m_a,\Ca)$ plane and receive important contributions from flavour observables, beam-dump experiments, and collider searches~\cite{Beacham:2019nyx,Garcia:2023awt}.
% A global recast of those limits is not attempted here.
% In particular, overlaying revised low-energy exclusions on contours obtained with the free-quark lifetime model would mix two different descriptions of the same benchmark.
% The projections below therefore isolate the ANUBIS reach of the specified production channels and state explicitly where the decay modelling limits their interpretation.\\

% \subsection{Fermionic decays and lifetime}
% \label{sec:decays}
For a charged lepton, the tree-level partial width following from Equation~\eqref{eq:pseudoscalar} is
\begin{equation}
\Gamma(a\to\ell^+\ell^-)=
\frac{\Ca^2m_a m_\ell^2}{8\pi\Lambda^2}
\sqrt{1-\frac{4m_\ell^2}{m_a^2}},
\label{eq:lepwidth}
\end{equation}
and at the parton level the corresponding quark width is
\begin{equation}
\Gamma(a\to q\overline q)=
\frac{N_c\Ca^2m_a m_q^2}{8\pi\Lambda^2}
\sqrt{1-\frac{4m_q^2}{m_a^2}},
\label{eq:quarkwidth}
\end{equation}
where $N_c=3$ is the colour factor.
At fixed mass, all partial widths therefore scale as $\Ca^2$, and the proper decay length scales as $c\tau_a\propto \Ca^{-2}$.
The lifetime can change abruptly with $m_a$ as a heavier fermion channel becomes kinematically accessible, because the partial widths include a factor of $m_f^2$.

Equations~\eqref{eq:lepwidth} and~\eqref{eq:quarkwidth} define the decay modes considered for BC10h in this article.
They are reliable for leptonic modes and provide a simple perturbative description at sufficiently high masses, but they are not a complete treatment of low-energy hadronic ALP decays.
Below a few \GeV, the physical final states are hadrons rather than free quarks, and chiral dynamics, ALP mixing with neutral pseudoscalar mesons, resonances, and renormalisation-group evolution can modify both the total width and the visible branching fractions~\cite{Garcia:2023awt}.
The sharp changes in sensitivity at nominal $2m_q$ thresholds presented in Section~\ref{sec:results_hloop} should consequently be regarded as features of the benchmark implementation rather than physical resonance predictions.
This limitation is stated explicitly rather than absorbed into an uncontrolled correction of the event samples.

%%%%%%%%%%%%%%%%%%%%%%%%%%%%%%%%%%%%%%%
\section{Analysis with SET-ANUBIS}
\label{sec:analysis}
%%%%%%%%%%%%%%%%%%%%%%%%%%%%%%%%%%%%%%%
\subsection{Signal generation}
\label{sec:generation}
The analysis uses \setanubis~v1.0.0, a modular pipeline for model configuration, event generation, decay handling, detector propagation, event selection, and sensitivity construction~\cite{Reymermier:SETANUBISv100,erner2026setanubismodularpipelineanubis}.
This ensures that changes in geometry or lifetime are propagated without separate analysis scripts or manual event reprocessing, allowing for consistent results across different models.

Signal samples are generated for proton--proton collisions at $\sqrt{s}=14~\TeV$ with \madversion{3.5.8}~\cite{Alwall:2014hca} and the NNPDF2.3 NLO set with $\alpha_s(m_Z)=0.118$~\cite{Ball:2012cx}.
The ALP interactions are implemented with the ALPsEFT UFO model~\cite{ALPsEFTUFO,Alloul:2013bka,Brivio:2017ije}.
For the Higgs channel in BC10h, the effective gluon-gluon fusion vertex is supplied by the Higgs effective theory model~\cite{HiggsEFTUFO}, while the $hZa$ interaction is implemented manually and normalised to Equations~\eqref{eq:ceff}--\eqref{eq:hxsec}.
ALP decays are handled with \madspin~\cite{Artoisenet:2012st}, and parton showering and hadronisation are performed with \pythia[~8]~\cite{Bierlich:2022pfr}.

For an event with boost $\beta\gamma$, the probability density for a decay at a distance $x$ from the interaction point is
\begin{equation}
P(x\mid\beta\gamma,c\tau_a)=
\frac{1}{\beta\gamma c\tau_a}
\exp\!\left[-\frac{x}{\beta\gamma c\tau_a}\right].
\label{eq:decaypdf}
\end{equation}
At fixed mass the event kinematics are independent of the overall gluon or fermion coupling, while the proper lifetime follows $c\tau_a\propto C_X^{-2}$, where $X$ is $\tilde{G}$ or $a\Phi$.
\setanubis therefore reuses each generated kinematic sample across the coupling grid by resampling the decay position with Equation~\eqref{eq:decaypdf}.
This removes the need to regenerate events at every lifetime hypothesis.

For the photophobic case, the 32 simulated masses are
$m_a\in\{$0.05, 0.07, 0.1, 0.13, 0.2, 0.3, 0.5, 0.7, 1, 1.3, 2, 3, 5, 7, 10, 12, 13, 15, 20, 30, 50, 55, 60, 75, 80, 85, 90, 100, 105, 110, 125, 150$\}$~\GeV.
The couplings considered cover ${10^{-11}\leq C_{\tilde{G}} \leq10^{-0.5}}$ in logarithmic steps of $0.1$ for a total of 106 couplings. For each $(m_a,C_{\tilde{G}})$ point, there are $\mathcal{O}(10^5)$ events from generated and reweighted samples.

For BC10h the simulated masses cover ${10^{-1.25}\leq m_a\leq10^{1.5}~\GeV}$
in logarithmic steps of $0.25$, giving 12 mass hypotheses. The couplings considered cover $10^{-6}\leq\Ca\leq10^{-2.2}$ in logarithmic steps of $0.1$, giving a total of 39 couplings.
Each mass and each kinematically open fermion decay mode has $\mathcal{O}(10^{4})$ events from generated and reweighted samples.
%$3\times10^4$ events are generated.
% The decay-mode samples are combined after selection with the branching fractions from Equations~\eqref{eq:lepwidth} and~\eqref{eq:quarkwidth}.

% For resonant production, the inclusive Higgs cross section and width are taken as
% \begin{equation}
% \sigma(pp\to h)=54.67~\pb,
% \qquad \Gamma_h^{\mathrm{SM}}=4.088~\MeV,
% \end{equation}
% for $m_h=125~\GeV$ at $\sqrt{s}=14~\TeV$~\cite{deFlorian:2016spz}.
% The generated Higgs samples are normalised to this cross section after applying the branching fraction in Equation~\eqref{eq:hbr}.

\subsection{ANUBIS acceptance and event selection}
\label{sec:selection}
The fiducial geometry follows the ceiling configuration of Ref.~\cite{ANUBIS:2025sgg}, using the \atlas coordinate system\footnote{%
\anubis adopts the right-handed coordinate system of \atlas, where its origin is at the nominal interaction point (IP) in the centre of the detector and the $z$-axis points along the beam pipe. The $x$-axis points from the IP to the centre of the LHC ring, and the $y$-axis points upwards.
Polar coordinates ($r$, $\phi$) are used in the transverse plane, $\phi$ being the azimuthal angle around the $z$-axis.
The pseudorapidity is defined in terms of the polar angle $\theta$ as $\eta \equiv-\ln \tan(\theta/2)$ and is equal to the rapidity $y = \frac12 \ln \left(\frac{E+p_z}{E-p_z}\right)$
in the relativistic limit.
Angular distance is measured in units of $\Delta R \equiv \sqrt{\Delta\eta^2+\Delta\phi^2}$.
}.
The sequential selections applied by \setanubis are summarised in Table~\ref{tab:Selection}.
The first stage requires the ALP trajectory and decay point to lie in the air-filled volume between \atlas and the tracking layers.
A displaced vertex is reconstructable when at least two charged daughters each intersect at least two tracking layers.
The momentum threshold ensures that the long-lived particle and its daughters are sufficiently energetic for charged-particle traversal of the RPC system.
The missing-transverse-momentum and isolation requirements target event topologies in which the LLP is separated from prompt QCD activity.
They also suppress neutral-hadron backgrounds associated with jets.

\begin{table*}[ht]
\caption{
Summary of the selection strategy implemented in the \setanubis framework to determine the geometric and kinematic acceptance for LLP decays.
The $(x,y,z)$ positions of the decay vertices are determined relative to the \atlas IP; $r_\text{CoC}$ represents the radius from the centre of curvature of the \atlas cavern ceiling ($x=-1.7~\metre$, $y=3.52~\metre$); and $r_\text{IP}$ is the radius from the \atlas IP~\cite{ANUBIS:HNLemu2026}.
%($x=1.7~\metre$, $y=-2.38~\metre$, $z=0~\metre$).
The selections are applied successively in the order shown.}
\label{tab:Selection}
\begin{tabular}{lll}
\hline
\textbf{Category} & \textbf{Description}      & \textbf{Requirement}
\\ \hline
\hline
\multicolumn{1}{l|}{\multirow{5}{*}{\textbf{Geometric acceptance}}} &
  \multirow{2}{*}{LLP decay within the \atlas cavern} &
  $-17.2<x<13.8~\metre$,  $y > -11.37~\metre$, $|z| < 26.4~\metre$,\\
\multicolumn{1}{c|}{} & & $ r_\text{CoC} < 20~\metre$ \\
\multicolumn{1}{c|}{} & \atlas veto               &  Not within ($|z|<22~\metre$ and $r_\text{IP} < 9.5~\metre$)                  \\
\multicolumn{1}{c|}{} & LLP-\anubis intersection       & $N_\text{inters.} \geq 2$ layers                     \\
\multicolumn{1}{c|}{} & LLP decay charged-track multiplicity              & $N_\text{tracks}\geq2$, $N_\text{hit} \geq 2$ layers per track \\ \hline
\multicolumn{1}{l|}{\multirow{4}{*}{\textbf{Kinematic acceptance}}} &
LLP momentum & $|\vec p_\text{LLP}| > 0.1$~\GeV\\
\multicolumn{1}{c|}{} & \met requirement               & $\met > 30~\GeV$                                \\
\multicolumn{1}{c|}{} & LLP isolation from jets              & $\Delta R_\text{LLP,jet} > 0.5$ for $p_\text{T}^\text{jet} > 15~\GeV$                           \\
\multicolumn{1}{c|}{} & LLP isolation from energetic tracks & $\Delta R_\text{LLP,trk} > 0.5$ for $p_\text{T}^\text{trk} > 5~\GeV$                       \\ \hline
\end{tabular}
\end{table*}

\begin{figure}[ht]
    \centering
    \includegraphics[width=0.94\linewidth]{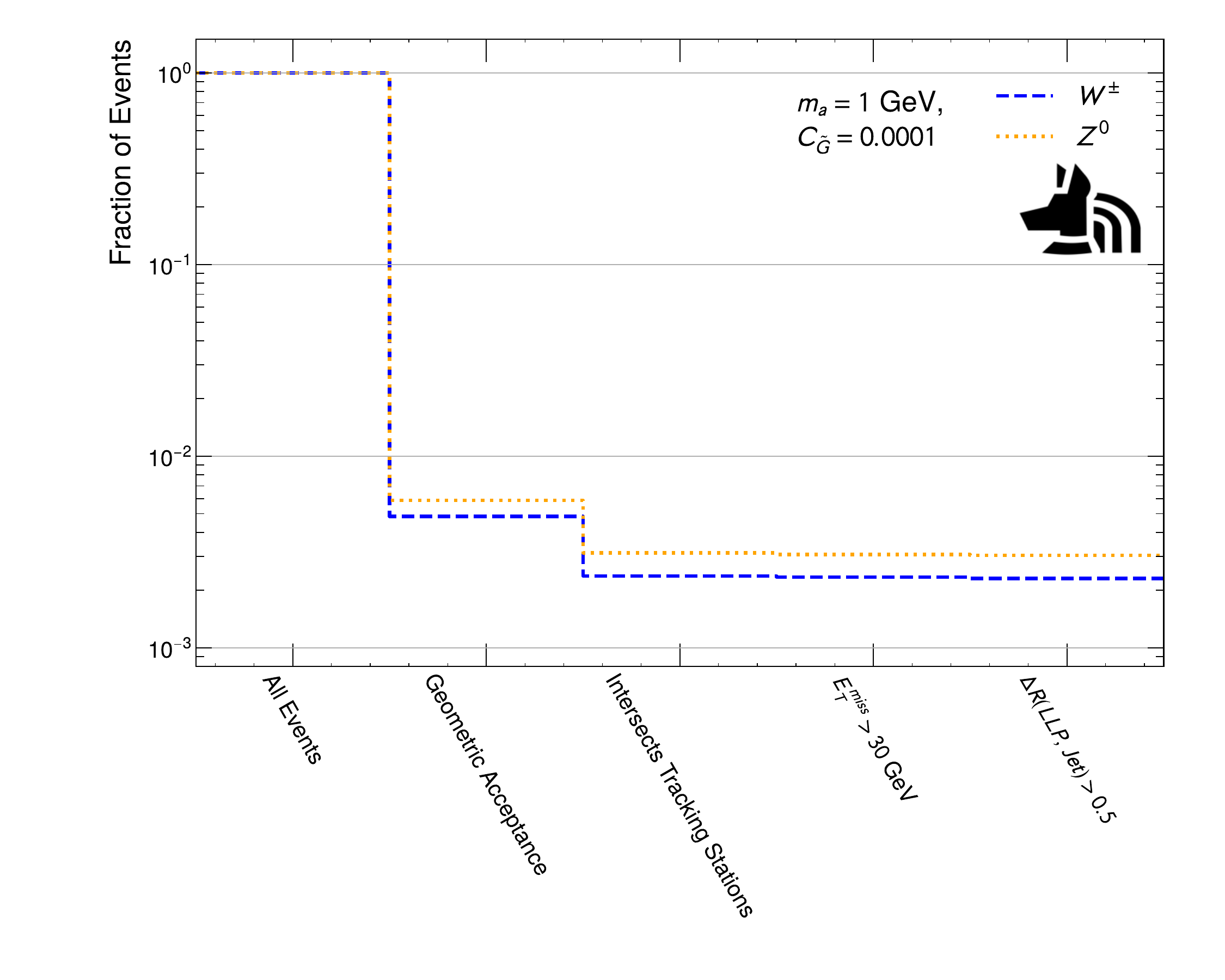}
    \includegraphics[width=0.94\linewidth]{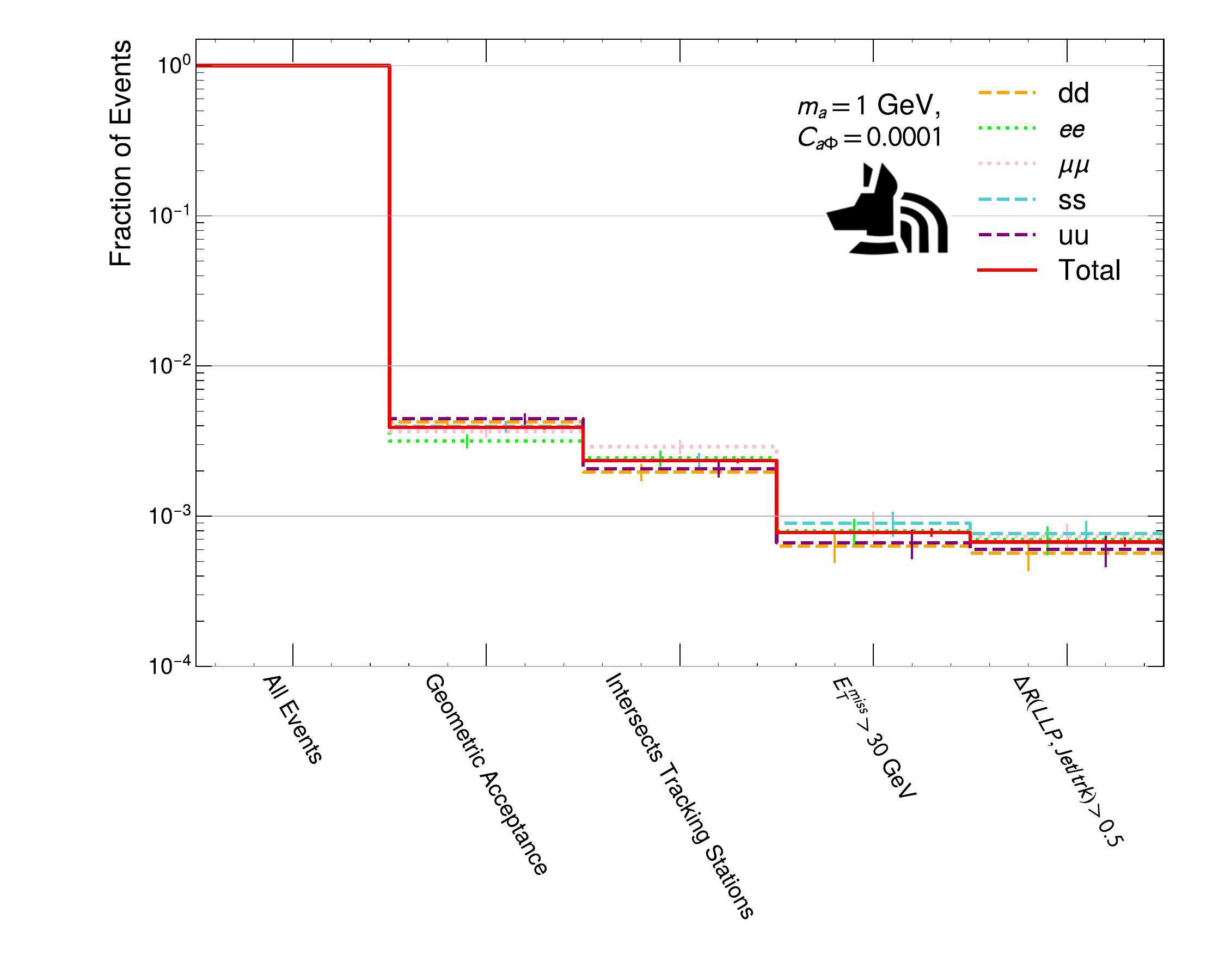}
    \caption{Fraction of surviving signal events after successively applying the selections, for the photophobic model with $W^{\pm}$ and $Z$ production (top) and for the BC10h model for each charged fermion decay mode (bottom), for a benchmark scenario with $m_a = 1~\GeV$ and a gluon or fermion coupling of $10^{-4}$.}
    \label{fig:Cutflow}
\end{figure}

The impact of applying the successive selections from Table~\ref{tab:Selection} is shown in Figure~\ref{fig:Cutflow} for $m_a=1~\GeV$ and a gluon (fermion) coupling of $10^{-4}$ in the photophobic (BC10h) scenario.
The geometric acceptance has the greatest single impact on the event yield.
The impact is similar for the $W^\pm a$ and $Za$ production modes in the photophobic model, and for each charged-fermion decay mode in the BC10h scenario, as expected from their similar kinematics.

\subsection{Projected LLP yield}
\label{sec:yield}

The expected number of selected LLP decays, summed over production modes $\pi$ and decay modes $\delta$, is
\begin{equation}
N_{\mathrm{LLP}}=
\mathcal L\,\varepsilon_{\mathrm{sig}}\BR(V\rightarrow \text{leptons})
\sum_{\pi,\delta} \sigma_\pi\BR_\delta\,
\mathcal A_{\pi,\delta},
\label{eq:nsig}
\end{equation}
where $\sigma_\pi$ is the production cross-section of the LLP in mode $\pi$, $\BR_\delta$ is the branching fraction of LLP decay mode $\delta$, and $\mathcal A_{\pi,\delta}$ is the channel-specific geometric and kinematic acceptance.
An integrated luminosity of $\mathcal L=3~\invab$ is considered, as expected for the HL-LHC and in line with the conventions of the transverse- and forward-detector communities.
A vertex-level signal efficiency of $\varepsilon_{\mathrm{sig}}=0.5$ is conservatively assumed after the geometric and kinematic selections, following the baseline \anubis projections~\cite{ANUBIS:2025sgg}.
% The finite-sample uncertainty on an acceptance obtained from $N_{\mathrm{gen}}$ events is treated as binomial,
% \begin{equation}
% \delta\mathcal A_{p,f}=
% \sqrt{\frac{\mathcal A_{p,f}(1-\mathcal A_{p,f})}{N_{\mathrm{gen}}}}.
% \label{eq:accerr}
% \end{equation}
% These uncertainties are propagated to the quoted event yields.
The factor $\BR(V\to\mathrm{leptons})$ accounts for the prompt vector boson decays into leptons. 
It is taken as ${\BR(W\to\mathrm{leptons})=0.327}$, counting decays to $\ell^-\bar\nu_\ell+\text{h.c.}$, and ${\BR(Z\to\mathrm{leptons})=0.101}$, counting decays to $\ell^+\ell^-$~\cite{PhysRevD.110.030001}.
For both ALP models, the 95\% confidence-level (CL) exclusion contour assumes an effectively background-free displaced-vertex search, corresponding to $N_{\mathrm{LLP}}=4$. 
The assumption of negligible background is justified by the leptonic decay of the prompt vector boson, which provides a clear signature in the \atlas detector and reduces the conservative background estimate to well below one event, as discussed in Ref.~\cite{ANUBIS:HNLemu2026}. 
The triggering on and reconstruction of the prompt vector boson are taken to be fully efficient at this stage.

% The exact Poisson upper limit for zero observed events and zero expected background is close to three events; the four-event convention is retained as a conservative allowance for unmodelled detector and background effects and for consistency with previous transverse-detector projections~\cite{Read:2002hq,ANUBIS:2025sgg,Bauer:2018uxu}.

%%%%%%%%%%%%%%%%%%%%%%%%%%%%%%%%%%%%%%%
\section{Results}
\label{sec:results}

\subsection{Photophobic model}
\label{sec:photophobicResults}

\begin{figure}[htb]
    \centering
    \includegraphics[width=0.95\linewidth]{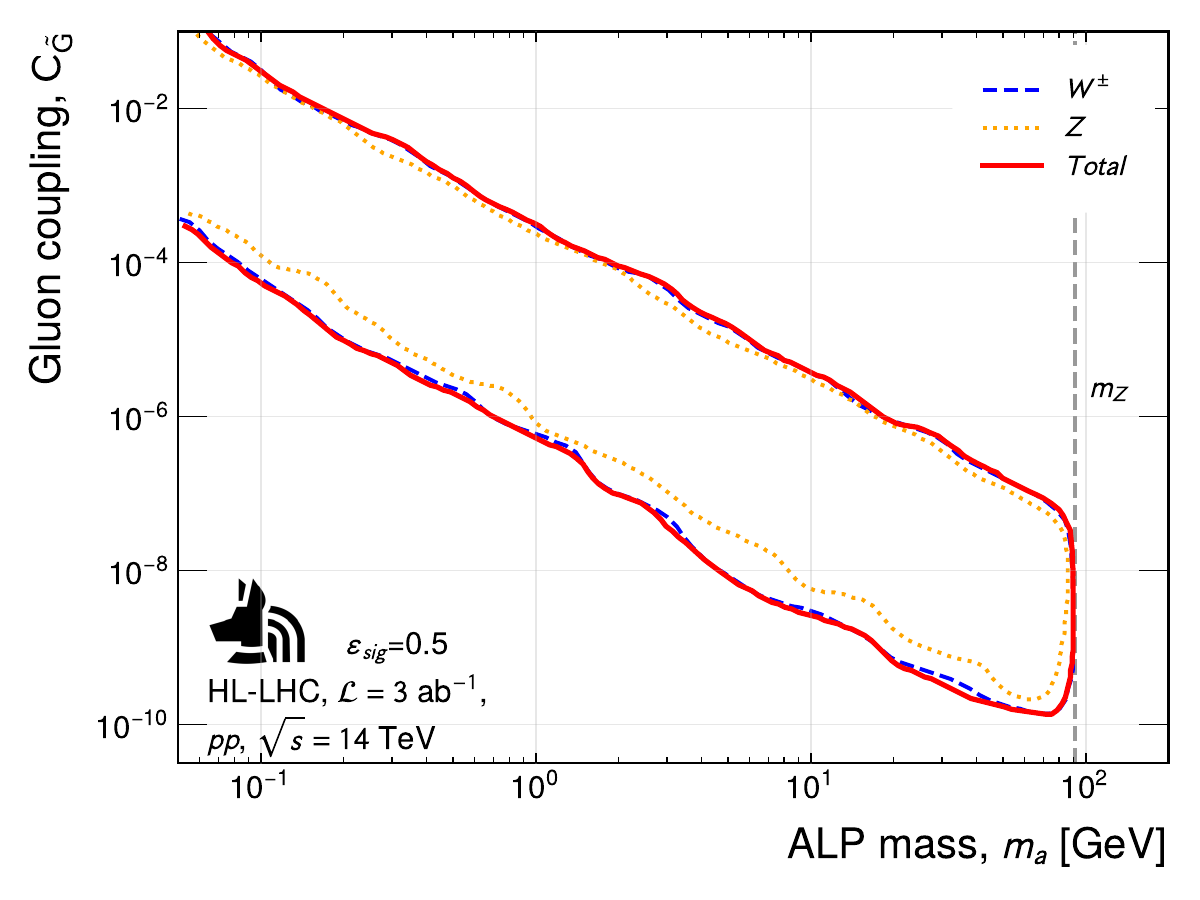}
    \caption{
    Projected exclusion sensitivity of the \anubis detector at 95\% CL to a photophobic ALP produced in association with a weak vector boson, as a function of the ALP mass, $m_a$, and the ALP--gluon coupling, $C_{\tilde{G}}$.
    The combined sensitivity~(solid line) is overlaid with the individual contributions from the $W^\pm$ and $Z$  production modes~(dashed and dotted lines, respectively).}
    \label{fig:contours_photophobic}
\end{figure}

The exclusion sensitivity contours for the production with an associated $W^\pm$ and $Z$ are shown in Figure~\ref{fig:contours_photophobic}, alongside the combination of the two. These contours run diagonally in the $(\ln m_a,\ln C_{\tilde{G}})$ plane, such that the smallest couplings are probed at the largest $m_a$.
This dependence is driven by the geometrical acceptance of the \anubis detector, which can be understood since $c\tau\propto C_{\tilde{G}}^{-2}$, while a smaller $m_a$ corresponds to a larger relativistic boost factor and hence a larger decay length, $\beta\gamma c\tau$. The peak sensitivity is $C_{\tilde{G}}=1.4\times10^{-10}$ at $m_a=71.4~\GeV$. 
The sensitivity is cut off at ${m_a\approx m_Z}$, where the decay $a\to Z\gamma$ becomes kinematically accessible; for small gluon couplings this mode dominates, and decays to gluons become negligible.

\begin{figure}[htb]
    \centering
    \includegraphics[width=0.95\linewidth]{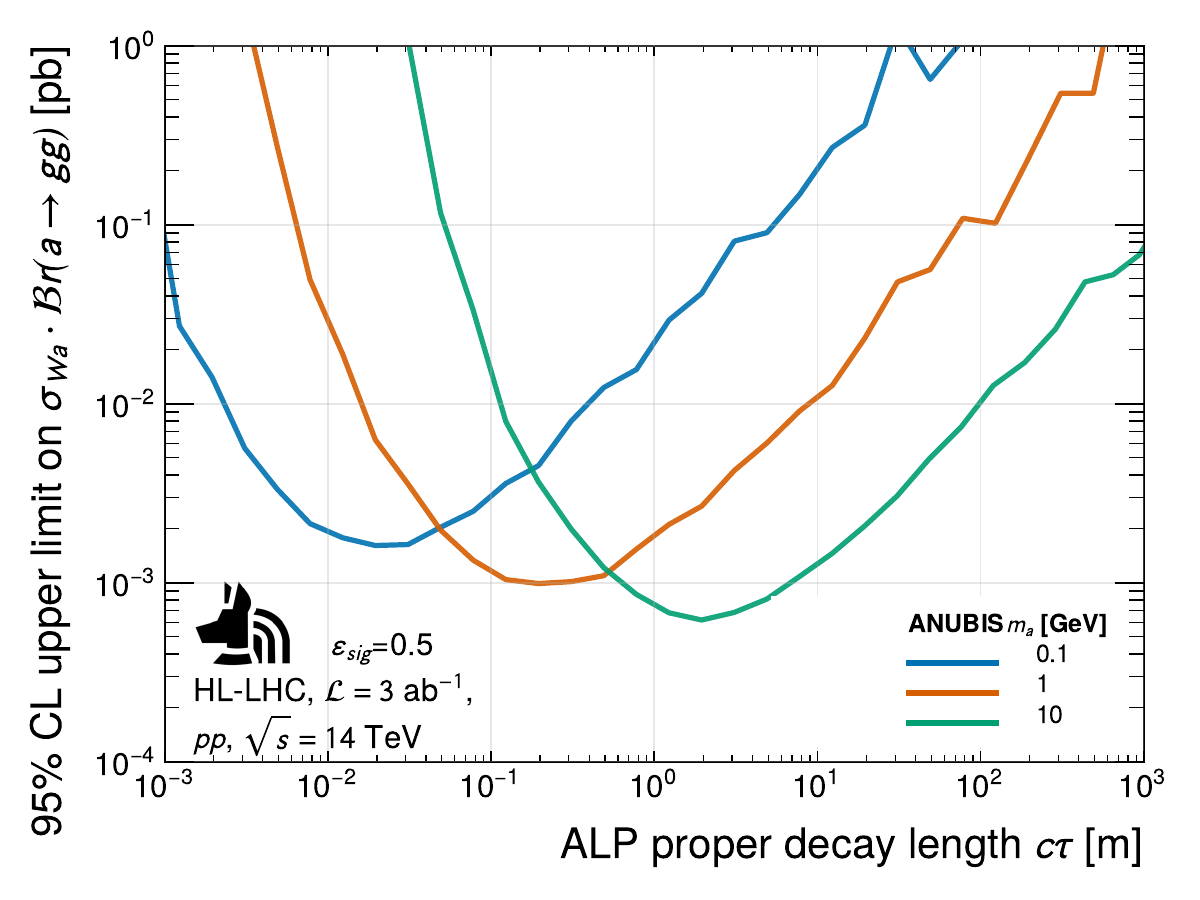}
    \includegraphics[width=0.95\linewidth]{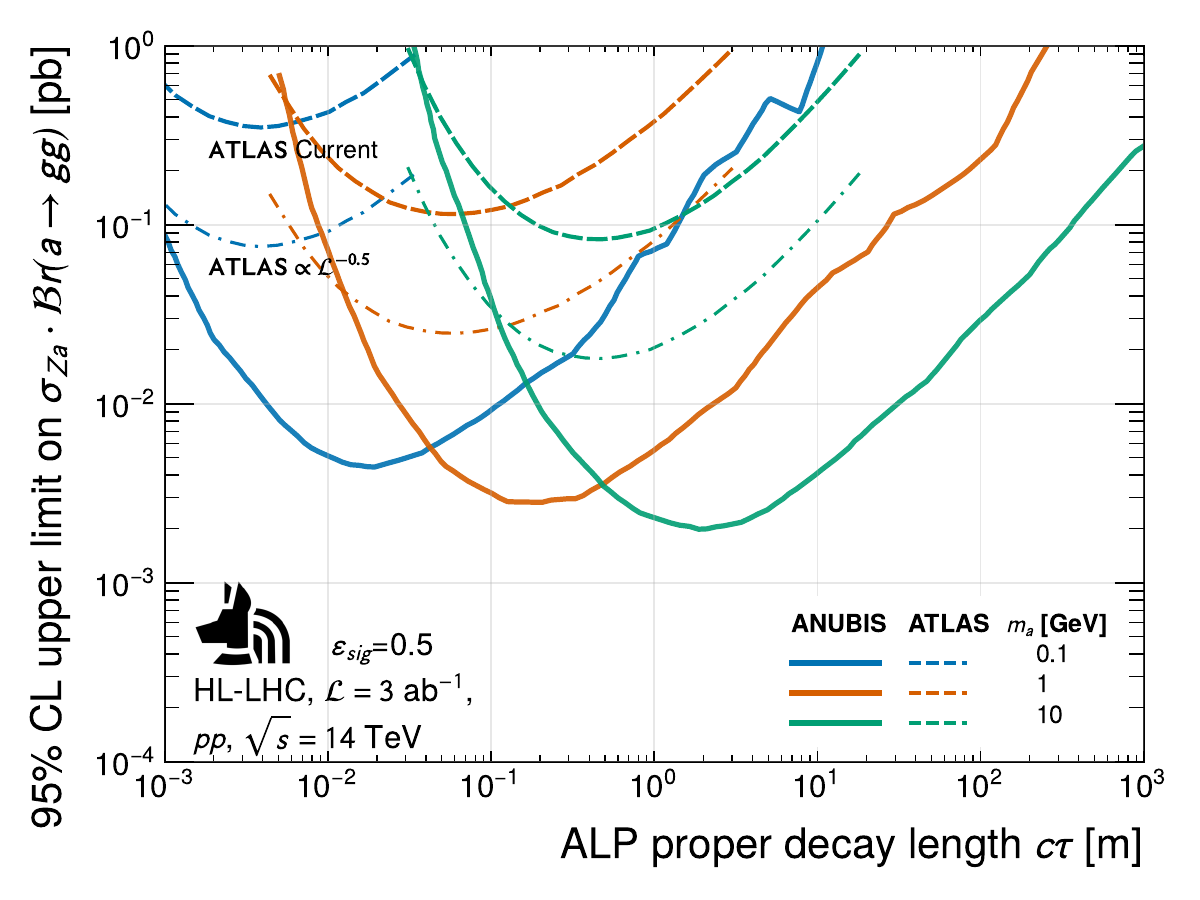}
    \caption{
    Projected exclusion sensitivity of the \anubis detector at 95\% CL on the product of the production cross-section and the ALP decay branching fraction, as a function of $c\tau$, for the photophobic ALP model. For reference, the existing exclusion limits from the \atlas experiment are shown in the $Za$ panel~\cite{ATLAS:2025pak}, scaled to the HL-LHC luminosity as described in the text.}
    \label{fig:crossSec_photophobic}
\end{figure}

The exclusion sensitivity for \anubis can also be considered in another parameter space, $c\tau$ versus ${\sigma_{Va}\cdot\BR(a\rightarrow gg)}$, which allows for flexible reinterpretation and direct comparison with experimental limits from \atlas~\cite{ATLAS:2025pak}. 
To this end, the gluon coupling is converted into $c\tau$ via Equation~\eqref{eq:GluonDecay} for a given $m_a$ value, while $\sigma_{Va}\cdot\BR(a\rightarrow gg)$ is obtained by rearranging Equation~\eqref{eq:nsig} and setting $N_{\mathrm{LLP}}=4$ explicitly. 
The latest results from \atlas consider ALPs with masses of 0.1, 1, 10, and $40~\GeV$ using 140~\invfb of $pp$ data at $\sqrt{s}=13~\TeV$~\cite{ATLAS:2025pak}. 
For a direct comparison with these \atlas limits, the \anubis sensitivity was evaluated for the same mass points, except for $m_a=40~\GeV$, whose experimental signature is similar to that for $m_a=10~\GeV$. 
The results are shown separately for an ALP produced in association with a $W^\pm$ or $Z$ boson in Figure~\ref{fig:crossSec_photophobic}. 
The fluctuations at larger $c\tau$ values are due to the limited size of the simulated samples in that region. 
The \atlas limits have also been scaled by $\sqrt{140~\invfb/3~\invab}\approx0.22$ to provide an approximate HL-LHC projection. 
This assumes that the number of background events, $b$, will approximately scale with the luminosity, $b\propto\mathcal{L}$, so that $\sigma_{Va}\cdot\BR(a\rightarrow gg)\propto\sqrt{b}/\mathcal{L}$.

Overall, the sensitivity of \anubis peaks at ${\sigma_{Va}\cdot\BR(a\rightarrow gg)=\mathcal{O}(10^{-3})}$~pb for both associated production modes and all ALP masses. 
The specific values are given in Table~\ref{tab:photophobicPeakSens}. 
The two production modes considered peak at the same $c\tau$ because of their comparable geometric acceptance; the difference in their peak sensitivity arises from the factor of about three between the leptonic branching fractions of the associated vector bosons used for triggering in \atlas. 
For $Za$ production, \anubis improves on the current \atlas exclusion limits by two orders of magnitude, and on the luminosity-scaled \atlas projection by one order of magnitude. 
Additionally, \anubis exceeds the $c\tau$ coverage of \atlas by approximately three orders of magnitude for several ALP masses.

\begin{table}[htb]
\caption{Peak exclusion sensitivities to $\sigma_{Va}\cdot\BR(a\rightarrow gg)$ for \anubis for different ALP masses alongside the associated $c\tau$ value.}
\label{tab:photophobicPeakSens}
\resizebox{\columnwidth}{!}{%
\begin{tabular}{l|ll|ll}
\hline
\multicolumn{1}{c|}{\multirow{2}{*}{$m_a$ [GeV]}} & \multicolumn{2}{c|}{$Wa$ Production}       & \multicolumn{2}{c}{$Za$ Production}       \\
\multicolumn{1}{c|}{}                            & $\sigma_{Wa}\cdot\BR(a\rightarrow gg)$ [pb]  & $c\tau$ [m] & $\sigma_{Za}\cdot\BR(a\rightarrow gg)$ [pb]  & $c\tau$ [m] \\ \hline
0.1 & $1.6\times10^{-3}$ & 0.02 & $4.4\times10^{-3}$ & 0.02\\
1   & $9.8\times10^{-4}$ & 0.20 & $2.8\times10^{-3}$ & 0.19\\
10  & $6.2\times10^{-4}$ & 1.95 & $2.0\times10^{-3}$ & 1.94 \\ \hline
\end{tabular}%
}
\end{table}
% ATLAS Z prod original (scaled)
% 0.1: 0.353(0.075) @ 0.004
% 1: 0.115(0.025) @ 0.06
% 10: 0.083(0.018) @ 0.47

\subsection{BC10 model}
\label{sec:results_bc10}
The `vanilla' BC10 model featuring a fermion-coupled ALP is studied in the $pp\rightarrow Va$ production channel where $a\rightarrow f\overline f$ and $V\in\{W^\pm,Z\}$. 
As for the photophobic model, the event yield of LLP decays inside the \anubis detector after all selections is determined, scanning over the free parameters of the model $(m_a,\Ca)$. 
With $3~\invab$, \anubis is found to have insufficient sensitivity to the vanilla BC10 scenario because of the small production rates, which yield at most $\order{1}$ event in the signal region.

\subsection{BC10h model}
\label{sec:results_hloop}
The large inclusive Higgs sample at the HL-LHC substantially increases the ALP production rate. 
In the strict fermion-coupled benchmark, the $pp\to h\to Za$ process is added to the vanilla BC10 model with the $h\to Za$ decay generated solely by the top-quark loop, \ie $\CZseven=0$ and $\CZh\simeq+0.016\Ca$. 
In this case, the peak expected \anubis yield remains just below the four-event threshold with 3~\invab.
%
% \begin{figure}[t]
% \centering
% \includegraphics[width=\columnwidth]{img/higgs_loop.pdf}
% \caption{Expected signal yield for fermion-coupled ALPs produced through $pp\to h\to Za$ when the $hZa$ amplitude is generated only by the fermion loop. The black curve is an illustrative $N_{\mathrm{sig}}=10^{-8}$ contour.}
% \label{fig:higgsloop}
% \end{figure}

\begin{figure}[t]
\centering
\includegraphics[width=\columnwidth]{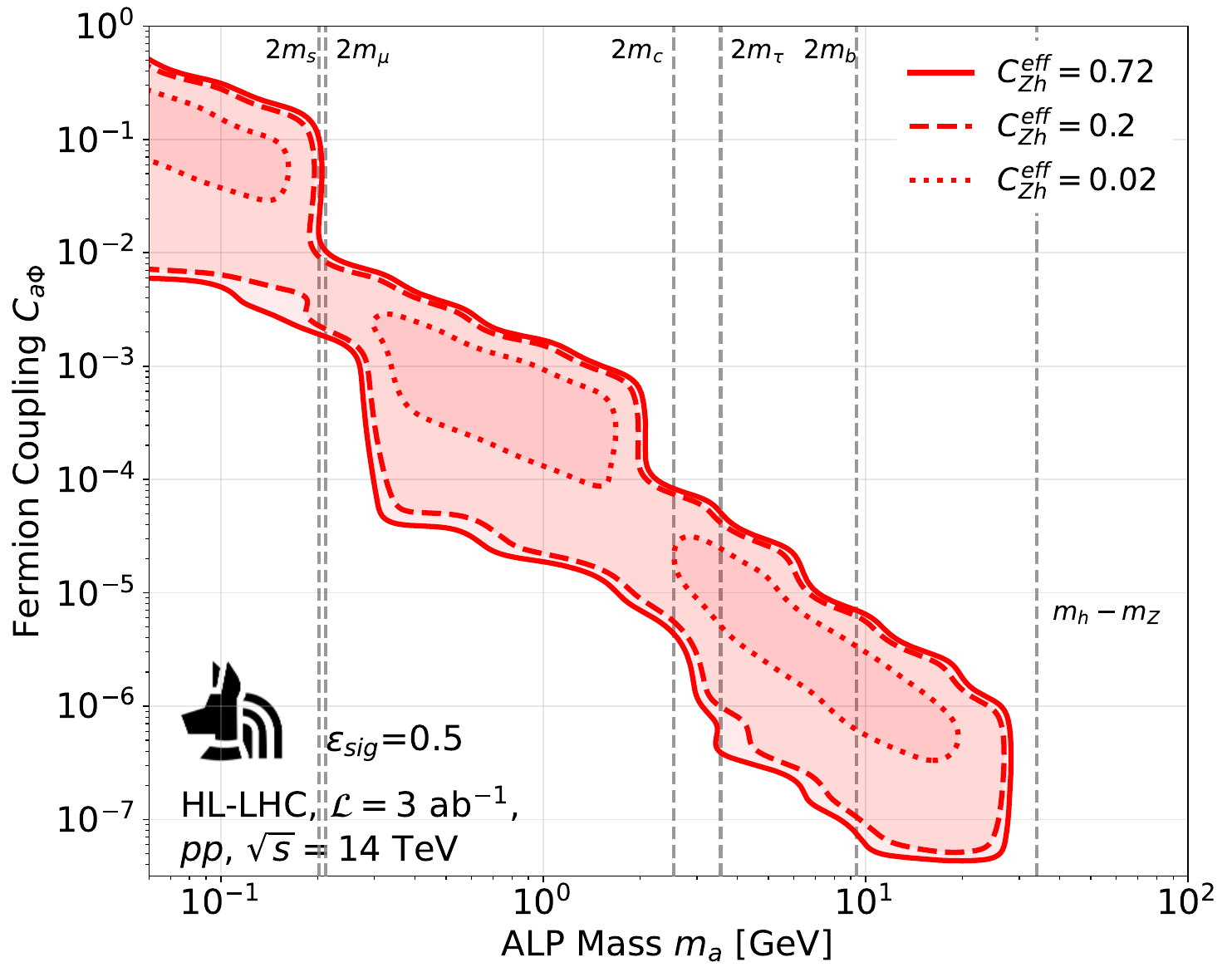}
\caption{
Projected exclusion sensitivity of the \anubis detector at 95\% CL to $pp\to h\to Za$ as a function of $m_a$ and the fermion coupling \Ca, for three values of the effective Higgs coefficient \CZh, which is varied independently of \Ca.
The solid, dashed, and dotted contours correspond to $\CZh=0.72$, $0.2$, and $0.02$, respectively.}
\label{fig:d7}
\end{figure}

% The maximum expected yield is
% \begin{equation}
% N_{\mathrm{sig}}=3.63\pm0.39
% \end{equation}
% at $m_a=10^{-1.25}~\GeV$ and $\Ca=10^{-0.5}$.
% The resonant cross section at this point is $1.58\times10^{-3}~\pb$, and the branching-ratio-weighted acceptance is $1.91\times10^{-3}$.
% The Higgs channel therefore increases the yield by roughly five orders of magnitude relative to $Za$ associated production.
% Nevertheless, the nominal central value remains below the conservative four-event boundary.
% The proximity of the highest-yield point to the threshold should not be read as a robust exclusion: it is sensitive to Monte Carlo statistics, the assumed signal efficiency, and the low-mass decay model.
% Within the stated benchmark assumptions, no closed 95\% confidence-level exclusion contour is obtained for a purely fermion-coupled ALP.
%
% \subsection{Independent Higgs--ALP interaction}
% \label{sec:results_d7}
The situation changes when BSM contributions to the effective Higgs coefficient in addition to the fermion loop are allowed, so that \CZh becomes a free parameter of the model.
Figure~\ref{fig:d7} shows the exclusion sensitivity contours in the $(m_a,\Ca)$ parameter space for $\CZh=0.72$, $0.2$, and $0.02$.
Here production is set by \CZh, while the displaced decay probability is controlled by \Ca. These values illustrate the sensitivity down to small values of \CZh; the largest, $\CZh=0.72$, corresponds to the legacy global constraint $\BR(h\to\mathrm{BSM})<0.34$ under the assumptions of Refs.~\cite{ATLAS:2016neq,Bauer:2017ris}.
Direct searches for $h\to Za$ with a hadronically decaying ALP currently constrain $\CZh/\Lambda$ to below 0.9--2~$\TeV^{-1}$ for $0.5<m_a<3.5~\GeV$~\cite{ATLAS:2024hZaHad}, which is weaker than this bound.
%OB   0.72 comes from the Run-1 ATLAS+CMS combination. The Run-2
%OB   combinations (Nature 607 (2022)) give Br_u ~ 0.12-0.16, i.e.
%OB   C_Zh^eff ~ 0.37. Checking Eq. (15) with Gamma_h^SM = 4.088 MeV:
%OB   Br = 0.34 gives 0.72 and Br = 0.12 gives ~0.37. 
%OB   It would be good to update the limit plot to 0.37...

For $\CZh=0.72$, the reach extends over a broad lifetime band from the low-mass boundary of the scan to the kinematic endpoint $m_a<m_h-m_Z$. The peak sensitivity is $\Ca=4.4\times10^{-8}$ for $m_a$ between 15.7 and $23.4~\GeV$.
Most of this band remains accessible at $\CZh=0.2$.
At $\CZh=0.02$, the reduced Higgs branching fraction moves both sides of the contour towards the region of largest geometric acceptance.
The overall diagonal shape of the band has the same origin as in the photophobic case, \ie the LLP boost. In addition, the contours show pronounced steps at the di-fermion thresholds, $m_a=2m_f$, where new decay channels become kinematically accessible.
% The upper and lower boundaries have different physical origins: the upper boundary corresponds to ALPs that decay before reaching ANUBIS, while the lower boundary is set by ALPs whose decay length is too large.

\begin{figure}[ht]
\centering
\includegraphics[width=0.95\columnwidth]{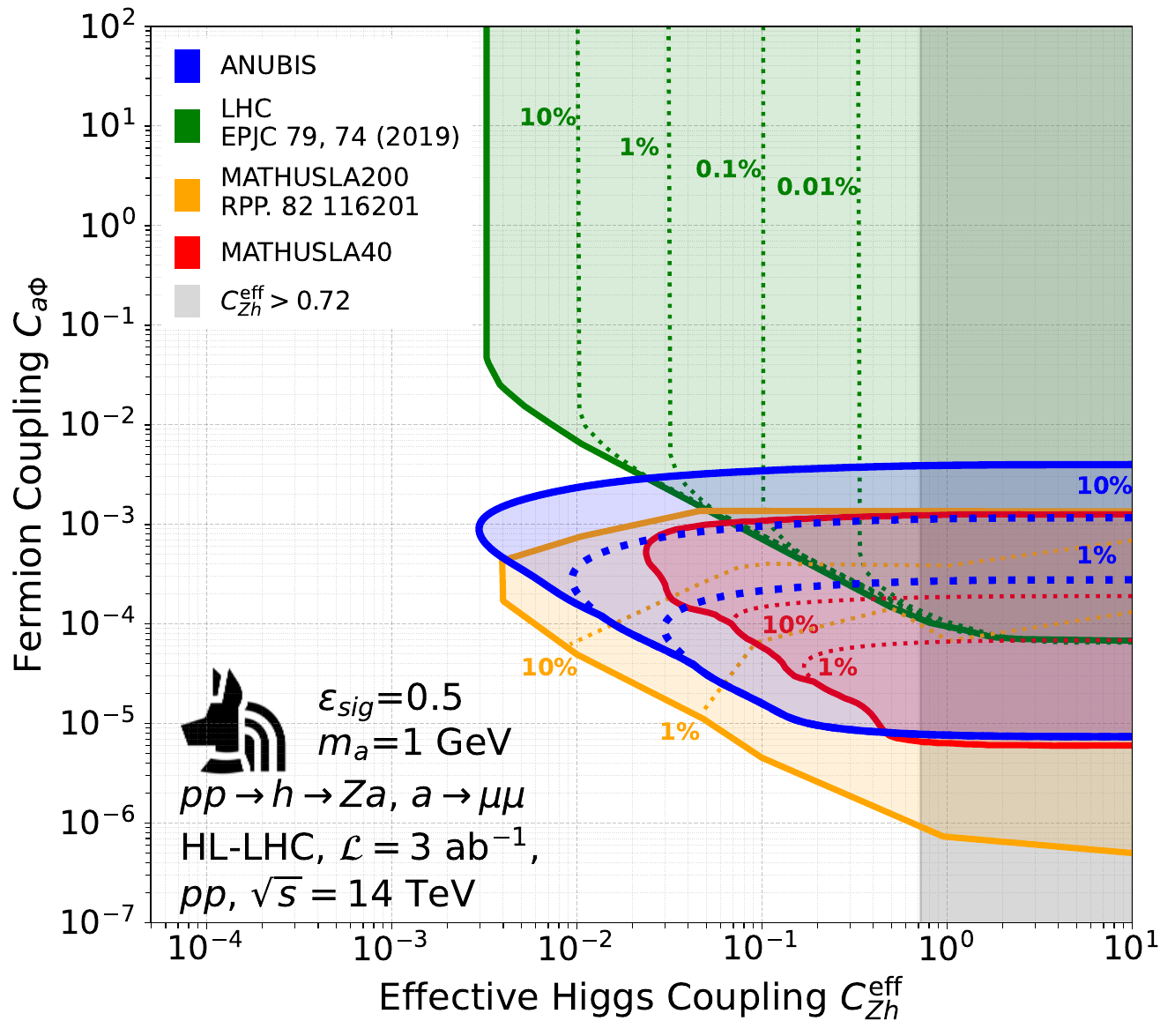}
\caption{
Projected exclusion sensitivity of the \anubis detector at 95\% CL as a function of \CZh and \Ca for $m_a=1~\GeV$ and $a\to\mu^+\mu^-$. The \anubis contour is compared with the projected limits for the main LHC detectors from Ref.~\cite{Bauer:2018uxu} and legacy MATHUSLA200 projections from Ref.~\cite{Curtin:2018mvb}.
%Limits for the current MATHUSLA40 geometry~\cite{MATHUSLA:2025strategy} have been derived by implementing the geometry within \setanubis and should be treated as an approximation to the sensitivity.
Solid contours assume a unit muonic branching fraction; decreasing branching fractions are indicated by labelled dotted curves. The grey region follows the exotic-Higgs-decay normalisation used for this benchmark comparison.}
\label{fig:comparison}
\end{figure}

% \subsection{Comparison with main-detector and MATHUSLA projections}
% \label{sec:comparison}
Since \CZh is a free parameter, the sensitivity can also be evaluated in the $(\CZh,\Ca)$ plane at fixed ALP mass. 
This plane was studied for \mathusla in Ref.~\cite{Curtin:2018mvb} for $m_a=1~\GeV$, assuming that the ALP decays exclusively to muons, $a\to\mu^+\mu^-$. 
For a direct comparison, a dedicated sample was simulated for this case following the strategy of Section~\ref{sec:generation}, but with $\BR(a\to\mu^+\mu^-)$ fixed to 1\%, 10\%, or 100\% in the generation and lifetime reweighting. 
The muonic branching fraction is treated as an independent phenomenological input. 
Figure~\ref{fig:comparison} shows the \anubis exclusion sensitivity in the $(\CZh,\Ca)$ parameter space in comparison to the existing projected limits from MATHUSLA200~\cite{Curtin:2018mvb} and LHC experiments~\cite{Bauer:2018uxu}. 
The \mathusla design has evolved since these projections were published and now foresees a smaller area of $40\times40~\metre^2$, referred to as MATHUSLA40~\cite{MATHUSLA:2025strategy}. 
To obtain MATHUSLA40 projections for comparison, a simplified geometry based on the dimensions given in Ref.~\cite{MATHUSLA:2025strategy} was implemented in \setanubis. 
The same simulated samples as for the \anubis projections were used, with only the geometric acceptance selection applied, and the resulting sensitivity for $N_{\mathrm{LLP}}=4$ is shown in Figure~\ref{fig:comparison}.
Given the approximations in this procedure, the MATHUSLA40 contour should be regarded as an approximate and conservative estimate of its sensitivity.
%It is not the branching fraction predicted by the universal partonic BC10 decay model at every point.

Figure~\ref{fig:comparison} shows that \anubis covers a significant region of unique sensitivity that cannot be probed by the main LHC experiments.
The main-detector benchmark is strongest at relatively large fermion couplings, where the ALP decays closer to the interaction point.
\mathusla reaches smaller effective Higgs couplings for part of the long-lifetime band because of its surface decay volume.
\anubis fills the intermediate region and extends to both shorter and longer lifetimes within the benchmark plane, reaching fermion couplings about forty times below the main-detector benchmark, while extending to values of \Ca about one order of magnitude larger and \CZh about a factor four smaller than the MATHUSLA40 approximation in parts of the plane. The overall area of coverage is similar for MATHUSLA200 and \anubis, but the projected MATHUSLA40 contour is entirely enclosed within that of \anubis.
% These ratios are approximate and depend on the common assumptions used to construct Figure~\ref{fig:comparison}, in particular the event thresholds, branching fractions, and reconstruction efficiencies.
Outside the excluded region of \CZh, \anubis reaches its peak sensitivity to the fermion coupling, $\Ca=7.9\times10^{-6}$, at $\CZh=0.72$, as expected from Figure~\ref{fig:d7}. 
The peak sensitivity to the effective Higgs coupling of $\CZh=3.0\times10^{-3}$ is reached at $\Ca=9.2\times10^{-4}$, which is comparable to the limit of the LHC main detectors and MATHUSLA200.
% The MATHUSLA40 boundary was cross-checked by matching the boosted decay length and signal multiplicity of the published $h\to XX$ benchmark to $h\to Za$.
% The validation is described in Appendix~\ref{app:mathusla}; the simplified simulation provides a conservative lower envelope and reproduces the qualitative placement of the reference contours.

\begin{figure}
    \centering
    \includegraphics[width=0.9\linewidth]{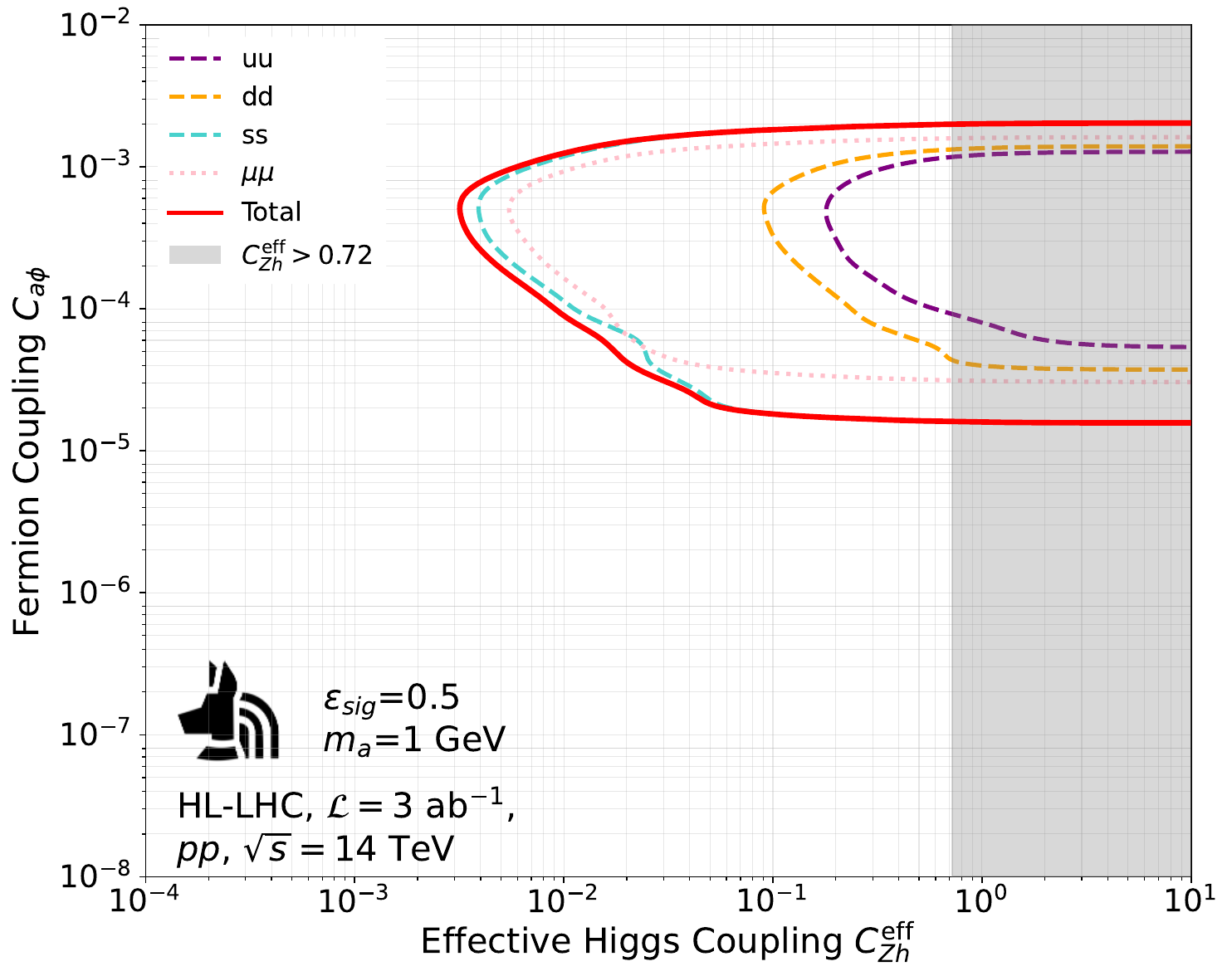}
    \includegraphics[width=0.9\linewidth]{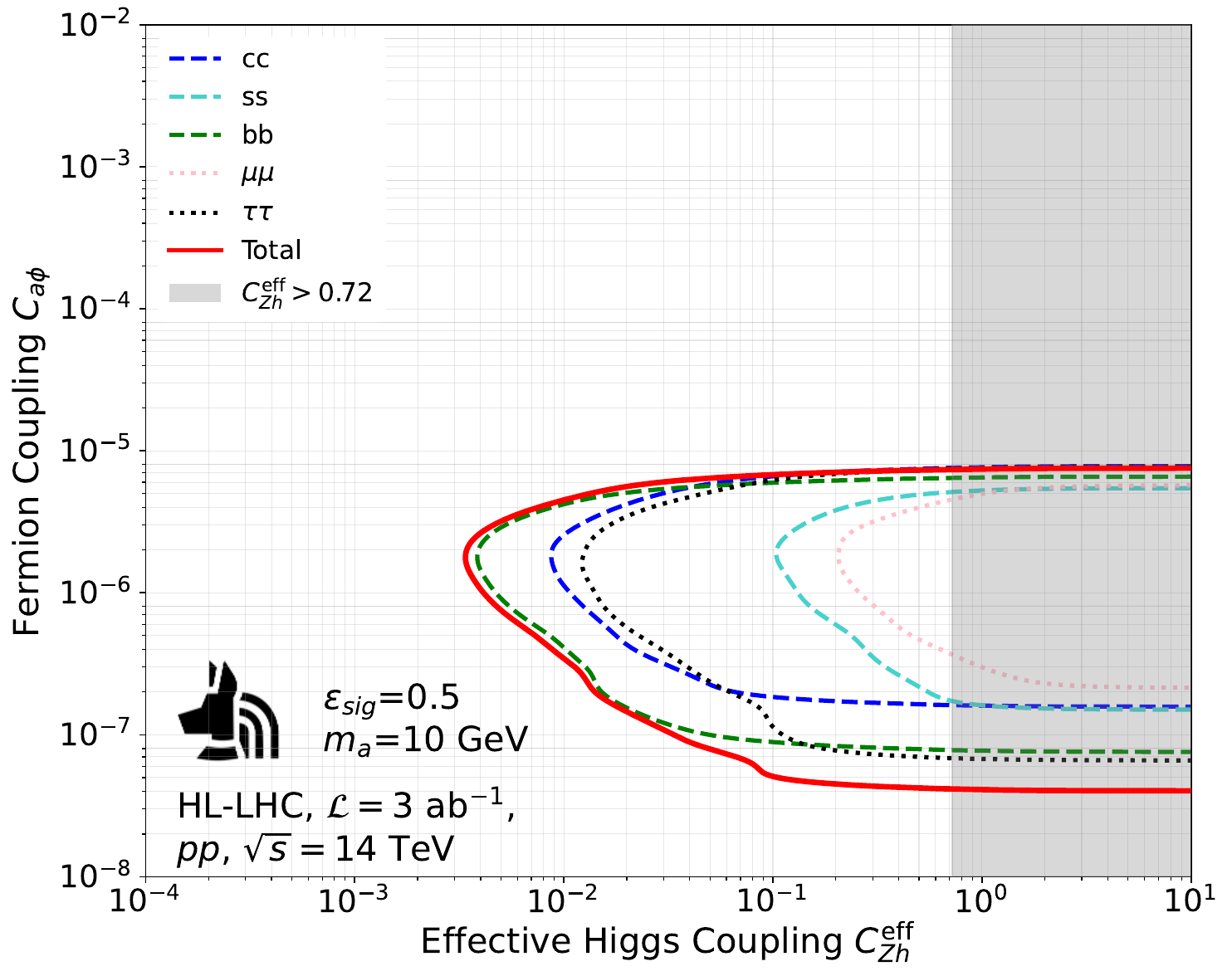}
    \caption{Projected exclusion sensitivity of the \anubis detector at 95\% CL as a function of \CZh and \Ca for (top) $m_a=1~\GeV$ and (bottom) $m_a=10~\GeV$ in the BC10h model, where the ALP decays to charged fermions.
    The contributions of the individual decay modes are shown as dashed and dotted lines, and their combination as a solid line.}
    \label{fig:BC10hANUBISfull}
\end{figure}

Instead of requiring $a\rightarrow\mu^+\mu^-$ only, all decay channels $a\rightarrow f\overline f$ can be considered, as in Figure~\ref{fig:d7}, to obtain a more realistic projection of the sensitivity of \anubis to the BC10h model. This is done in Figure~\ref{fig:BC10hANUBISfull} for two ALP masses, $m_a=1~\GeV$ and $m_a=10~\GeV$. 
The peak sensitivity to the fermion coupling outside the excluded \CZh region is $\Ca=1.6\times10^{-5}~(4.2\times10^{-8})$ at $\CZh=0.72$. 
For the effective Higgs coupling, the peak sensitivity is $\CZh=3.9\times10^{-3}~(3.4\times10^{-3})$ at $\Ca=5.4\times10^{-4}~(1.8\times10^{-4})$ for $m_a=1~(10)~\GeV$. 
The region of \Ca covered differs substantially between the two ALP masses, reflecting the mass dependence of the open decay channels and of their partial widths.

%%%%%%%%%%%%%%%%%%%%%%%%%%%%%%%%%%%%%%%
\section{Conclusions}
\label{sec:conclusion}
%%%%%%%%%%%%%%%%%%%%%%%%%%%%%%%%%%%%%%%
The projected sensitivity of the \anubis detector to photophobic and fermion-coupled ALPs has been evaluated for production in association with a weak vector boson in both models and, for the fermion-coupled model, additionally via the exotic Higgs decay $h\to Za$.
The analysis uses the \setanubis~v1.0.0 framework for signal generation, decay handling, detector propagation, selection, lifetime reweighting, and contour extraction.

The \anubis projections improve on the existing \atlas limits on $\sigma_{Va}\cdot\BR(a\rightarrow gg)$ for photophobic ALPs by two orders of magnitude and extend the $c\tau$ coverage by three orders of magnitude across the range of ALP masses studied. 
The sensitivity peaks at ${\sigma_{Va}\cdot\BR(a\rightarrow gg)=\mathcal{O}(10^{-3})~\pb}$ for all ALP masses considered. 
The \anubis detector is projected to reach a peak sensitivity to the gluon coupling of ${C_{\tilde{G}}=1.4\times10^{-10}}$ at $m_a=71.4~\GeV$. 

No sensitivity was found to the BC11 pure gluon-coupled ALP model in $Va$ final states.

% Direct production in association with a $W^\pm$ or $Z$ boson does not lead to an observable signal in the universal fermion-coupling benchmark.
% The mass-proportional ALP interaction suppresses emission from the light quarks that dominate the proton luminosity.
% The largest expected yields are of order $10^{-1}$ events for $W^\pm a$ and $10^{-4}$ events for $Za$, well below the four-event projection threshold.

% Resonant Higgs production largely removes the rate limitation.
% In strict BC10, the $hZa$ amplitude is generated by the top-quark loop and is tied to the same coupling that controls the ALP lifetime.
% The maximum expected yield is $3.63\pm0.39$ events, close to but below the nominal threshold.
% This result gives a quantitative boundary: the pure fermion benchmark is not excluded under the baseline assumptions, but modest changes in signal efficiency, production normalisation, or low-mass decay modelling can matter at the highest-yield point.

In the vanilla BC10 fermion-coupled ALP model, the peak expected yield in the \anubis detector remains just below the four-event threshold with 3~\invab. 
%No sensitivity was observed to the vanilla BC10 fermion-coupled ALP model. 
In the extended `BC10h' model, which includes an additional $hZa$ vertex, the \anubis detector is projected to cover a broad band in both the $(m_a,\Ca)$ and the $(\CZh,\Ca)$ planes.
% In the two-coupling interpretation, the Higgs coefficient fixes the production rate and the fermion coefficient fixes the decay length.
Considering only $a\to\mu^+\mu^-$ decays at $m_a=1~\GeV$, the \anubis detector uniquely probes a significant region of the parameter space that is complementary to the main LHC detector searches, especially in the transition between prompt and very long-lived decays. 
Considering the full set of fermionic ALP decays for $m_a=1$ and $10~\GeV$ yields peak sensitivities of $\Ca=1.6\times10^{-5}~(4.2\times10^{-8})$ at $\CZh=0.72$ and $\CZh=3.9\times10^{-3}~(3.4\times10^{-3})$ at $\Ca=5.4\times10^{-4}~(1.8\times10^{-4})$ for $m_a=1~(10)~\GeV$.

The sensitivity to all models considered in this paper is expected to improve significantly by exploiting ALP production mechanisms with nearby hadronic activity, which are currently rejected by the LLP isolation requirements.
This promising strategy would invalidate the assumptions of the current data-driven background estimate and would therefore require a dedicated background study, which is left for future study.

Overall, \anubis will significantly extend the coverage of the existing LHC experiments for several ALP models, in a region that cannot be covered by forward and fixed-target LLP detectors.

% $\bullet$ Our limit :
% \newline

% Free quark treatment of ALP decay (I suppose ?). leptonic modes, chiral hadronic amplitudes, pseudoscalar mixing, and perturbative QCD needed in low mass region.
% \newline

% channel-dependent prompt-object efficiencies and a dedicated background estimate using combined ATLAS--ANUBIS information instead of only four-event convention only?

%%%%%%%%%%%%%%%%%%%%%%%%%%%%%%%%%%%%%%%
%\section*{Acknowledgements}
%%%%%%%%%%%%%%%%%%%%%%%%%%%%%%%%%%%%%%%
\begin{acknowledgments} 
We thank everyone who contributed to the development and validation of the \setanubis framework, and the members of the \anubis Collaboration who contributed to the detector geometry and analysis strategy.
We gratefully acknowledge the support of UKRI through the Future Leaders Fellowship scheme.
\end{acknowledgments}

\clearpage
\bibliographystyle{JHEP}
\bibliography{bibs/alp,bibs/theory,bibs/general,bibs/anubis-standard-refs}

@article{Brivio:2017ije,
    author = "Brivio, Ilaria and Gavela, M. B. and Merlo, Luca and Mimasu, Ken and No, Jose M. and del Rey, Roberto and Sanz, Veronica",
    title = "{ALPs Effective Field Theory and Collider Signatures}",
    eprint = "1701.05379",
    archivePrefix = "arXiv",
    primaryClass = "hep-ph",
    doi = "10.1140/epjc/s10052-017-5111-3",
    journal = "Eur. Phys. J. C",
    volume = "77",
    number = "8",
    pages = "572",
    year = "2017"
}

@article{Bauer:2018uxu,
    author = "Bauer, Martin and Heiles, Mathias and Neubert, Matthias and Thamm, Andrea",
    title = "{Axion-Like Particles at Future Colliders}",
    eprint = "1808.10323",
    archivePrefix = "arXiv",
    primaryClass = "hep-ph",
    doi = "10.1140/epjc/s10052-019-6587-9",
    journal = "Eur. Phys. J. C",
    volume = "79",
    number = "1",
    pages = "74",
    year = "2019"
}

@article{Garcia:2023awt,
    author = "Dalla Valle Garcia, Giovani and Kahlhoefer, Felix and Ovchynnikov, Maksym and Zaporozhchenko, Andrii",
    title = "{Phenomenology of axion-like particles with universal fermion couplings -- revisited}",
    eprint = "2310.03524",
    archivePrefix = "arXiv",
    primaryClass = "hep-ph",
    doi = "10.1103/PhysRevD.109.055042",
    journal = "Phys. Rev. D",
    volume = "109",
    number = "5",
    pages = "055042",
    year = "2024"
}

@article{Choi:2020rgn,
    author = "Choi, Kiwoon and Im, Sang Hui and Shin, Chang Sub",
    title = "{Recent Progress in the Physics of Axions and Axion-Like Particles}",
    eprint = "2012.05029",
    archivePrefix = "arXiv",
    primaryClass = "hep-ph",
    doi = "10.1146/annurev-nucl-120720-031147",
    journal = "Ann. Rev. Nucl. Part. Sci.",
    volume = "71",
    pages = "225--252",
    year = "2021"
}

@article{Ball:2012cx,
    author = "Ball, Richard D. and others",
    title = "{Parton distributions with LHC data}",
    eprint = "1207.1303",
    archivePrefix = "arXiv",
    primaryClass = "hep-ph",
    doi = "10.1016/j.nuclphysb.2012.10.003",
    journal = "Nucl. Phys. B",
    volume = "867",
    pages = "244--289",
    year = "2013"
}

@article{ATLAS:2016neq,
    author = "Aad, Georges and others",
    collaboration = "ATLAS, CMS",
    title = "{Measurements of the Higgs boson production and decay rates and constraints on its couplings from a combined ATLAS and CMS analysis of the LHC pp collision data at sqrt(s) = 7 and 8 TeV}",
    eprint = "1606.02266",
    archivePrefix = "arXiv",
    primaryClass = "hep-ex",
    doi = "10.1007/JHEP08(2016)045",
    journal = "JHEP",
    volume = "08",
    pages = "045",
    year = "2016"
}

@misc{ALPsEFTUFO,
    author = "Brivio, Ilaria and others",
    title = "{ALPsEFT: Effective theories for a light axion-like particle}",
    howpublished = "FeynRules model database",
    year = "2017",
    note = "Model implementation used for event generation"
}

@misc{HiggsEFTUFO,
    author = "Christensen, Neil and Duhr, Claude",
    title = "{The Higgs effective theory}",
    howpublished = "FeynRules model database",
    year = "2011"
}

@article{MATHUSLA:2025strategy,
    author = "Aitken, Branden and others",
    collaboration = "MATHUSLA",
    title = "{MATHUSLA: An External Long-Lived Particle Detector to Maximize the Discovery Potential of the HL-LHC}",
    eprint = "2504.01999",
    archivePrefix = "arXiv",
    primaryClass = "physics.ins-det",
    month = "4",
    year = "2025"
}

@article{Aiko:2024xiv,
    author = "Aiko, Masashi and Endo, Motoi and Fridell, K{\aa}re",
    title = "{Heavy photophobic ALP at the LHC}",
    eprint = "2401.13323",
    archivePrefix = "arXiv",
    primaryClass = "hep-ph",
    doi = "10.1007/JHEP06(2024)194",
    journal = "JHEP",
    volume = "06",
    pages = "194",
    year = "2024"
}

@article{ATLAS:2020hZaHad,
    author = "Aad, Georges and others",
    collaboration = "ATLAS",
    title = "{Search for Higgs boson decays into a $Z$ boson and a light hadronically decaying resonance using 13 TeV $pp$ collision data from the ATLAS detector}",
    eprint = "2004.01678",
    archivePrefix = "arXiv",
    primaryClass = "hep-ex",
    doi = "10.1103/PhysRevLett.125.221802",
    journal = "Phys. Rev. Lett.",
    volume = "125",
    pages = "221802",
    year = "2020"
}

@article{ATLAS:2024hZaHad,
    author = "Aad, Georges and others",
    collaboration = "ATLAS",
    title = "{Search for Higgs boson decays into a $Z$ boson and a light hadronically decaying resonance in $pp$ collisions at $\sqrt{s}=13$ TeV with the ATLAS detector}",
    eprint = "2411.16361",
    archivePrefix = "arXiv",
    primaryClass = "hep-ex",
    doi = "10.1016/j.physletb.2025.139671",
    journal = "Phys. Lett. B",
    volume = "868",
    pages = "139671",
    year = "2025"
}

@article{ATLAS:2021H4l,
    author = "Aad, Georges and others",
    collaboration = "ATLAS",
    title = "{Search for Higgs bosons decaying into new spin-0 or spin-1 particles in four-lepton final states with the ATLAS detector with 139 fb$^{-1}$ of $pp$ collision data at $\sqrt{s}=13$ TeV}",
    eprint = "2110.13673",
    archivePrefix = "arXiv",
    primaryClass = "hep-ex",
    doi = "10.1007/JHEP03(2022)041",
    journal = "JHEP",
    volume = "03",
    pages = "041",
    year = "2022"
}

@article{bib:calratio,
    author = "Aad, Georges and others",
    collaboration = "ATLAS",
    title = "{Search for neutral long-lived particles that decay into displaced jets in the ATLAS calorimeter in association with leptons or jets using $pp$ collisions at $\sqrt{s}=13$ TeV}",
    eprint = "2407.09183",
    archivePrefix = "arXiv",
    primaryClass = "hep-ex",
    doi = "10.1007/JHEP11(2024)036",
    journal = "JHEP",
    volume = "11",
    pages = "036",
    year = "2024"
}

@article{bib:hdmi,
    author = "Aad, Georges and others",
    collaboration = "ATLAS",
    title = "{Search for light long-lived particles in $pp$ collisions at $\sqrt{s}=13$ TeV using displaced vertices in the ATLAS inner detector}",
    eprint = "2403.15332",
    archivePrefix = "arXiv",
    primaryClass = "hep-ex",
    doi = "10.1103/PhysRevLett.133.161803",
    journal = "Phys. Rev. Lett.",
    volume = "133",
    pages = "161803",
    year = "2024"
}

@article{ATLAS:2020lbl,
    author = "Aad, Georges and others",
    collaboration = "ATLAS",
    title = "{Measurement of light-by-light scattering and search for axion-like particles with 2.2 nb$^{-1}$ of Pb+Pb data with the ATLAS detector}",
    eprint = "2008.05355",
    archivePrefix = "arXiv",
    primaryClass = "hep-ex",
    doi = "10.1007/JHEP03(2021)243",
    journal = "JHEP",
    volume = "03",
    pages = "243",
    year = "2021"
}

@article{CMS:2018erd,
    author = "Sirunyan, Albert M and others",
    collaboration = "CMS",
    title = "{Evidence for light-by-light scattering and searches for axion-like particles in ultraperipheral PbPb collisions at $\sqrt{s_{\mathrm{NN}}}=5.02$ TeV}",
    eprint = "1810.04602",
    archivePrefix = "arXiv",
    primaryClass = "hep-ex",
    doi = "10.1016/j.physletb.2019.134826",
    journal = "Phys. Lett. B",
    volume = "797",
    pages = "134826",
    year = "2019"
}

@article{CMS:2022zfd,
    author = "Tumasyan, Armen and others",
    collaboration = "CMS, TOTEM",
    title = "{First search for exclusive diphoton production at high mass with tagged protons in proton-proton collisions at $\sqrt{s}=13$ TeV}",
    eprint = "2110.05916",
    archivePrefix = "arXiv",
    primaryClass = "hep-ex",
    doi = "10.1103/PhysRevLett.129.011801",
    journal = "Phys. Rev. Lett.",
    volume = "129",
    pages = "011801",
    year = "2022"
}

@article{ATLAS:2023etl,
    author = "Aad, Georges and others",
    collaboration = "ATLAS",
    title = "{Search for the decay of the Higgs boson to a $Z$ boson and a light pseudoscalar particle decaying to two photons}",
    eprint = "2312.01942",
    archivePrefix = "arXiv",
    primaryClass = "hep-ex",
    doi = "10.1016/j.physletb.2024.138536",
    journal = "Phys. Lett. B",
    volume = "850",
    pages = "138536",
    year = "2024"
}

@article{ATLAS:2023ian,
    author = "Aad, Georges and others",
    collaboration = "ATLAS",
    title = "{Search for short- and long-lived axion-like particles in $H\rightarrow aa\rightarrow4\gamma$ decays with the ATLAS experiment at the LHC}",
    eprint = "2312.03306",
    archivePrefix = "arXiv",
    primaryClass = "hep-ex",
    doi = "10.1140/epjc/s10052-024-12979-0",
    journal = "Eur. Phys. J. C",
    volume = "84",
    pages = "742",
    year = "2024"
}

@article{Bauer:2019vqk,
    author = "Bauer, Martin and Brandt, Oleg and Lee, Lawrence and Ohm, Christian",
    title = "{ANUBIS: Proposal to search for long-lived neutral particles in CERN service shafts}",
    journal = "accepted by APS Open Sci.",
    doi = "10.1103/n652-fgyb",
    eprint = "1909.13022",
    archivePrefix = "arXiv",
    primaryClass = "physics.ins-det",
    year = "2026"
}

@article{ANUBIS:2025sgg,
    author = "Adolphus, T and others",
    collaboration = "ANUBIS",
    title = "{The ANUBIS detector and its sensitivity to neutral long-lived particles}",
    journal = "accepted by Phys. Rev. D",
    doi = "10.1103/kd5j-d4fv",
    archivePrefix = "arXiv",
    primaryClass = "hep-ex",
    year = "2026",
}

@article{Curtin:2018mvb,
    author = "Curtin, David and others",
    title = "{Long-Lived Particles at the Energy Frontier: The MATHUSLA Physics Case}",
    eprint = "1806.07396",
    archivePrefix = "arXiv",
    primaryClass = "hep-ph",
    reportNumber = "FERMILAB-PUB-18-264-T",
    doi = "10.1088/1361-6633/ab28d6",
    journal = "Rept. Prog. Phys.",
    volume = "82",
    number = "11",
    pages = "116201",
    year = "2019"
}

@article{Alemany:2019vsk,
    author = "Alemany, R. and others",
    collaboration = "{PBC}",
    editor = "Jaeckel, J. and Lamont, M. and Vall\'ee, C.",
    title = "{Summary Report of Physics Beyond Colliders at CERN}",
    eprint = "1902.00260",
    archivePrefix = "arXiv",
    primaryClass = "hep-ex",
    reportNumber = "CERN-PBC-REPORT-2018-003",
    month = "2",
    year = "2019"
}

@article{Antel:2023hkf,
    author = "Antel, C. and others",
    title = "{Feebly-interacting particles: FIPs 2022 Workshop Report}",
    eprint = "2305.01715",
    archivePrefix = "arXiv",
    primaryClass = "hep-ph",
    reportNumber = "CERN-TH-2023-061, DESY-23-050, FERMILAB-PUB-23-149-PPD, INFN-23-14-LNF, JLAB-PHY-23-3789, LA-UR-23-21432, MITP-23-015",
    doi = "10.1140/epjc/s10052-023-12168-5",
    journal = "Eur. Phys. J. C",
    volume = "83",
    number = "12",
    pages = "1122",
    year = "2023"
}

@article{PhysRevD.110.030001,
  title = {Review of Particle Physics},
  author = {Navas, S. and others},
  collaboration = {Particle Data Group},
  journal = {Phys. Rev. D},
  volume = {110},
  issue = {3},
  pages = {030001},
  numpages = {5},
  year = {2024},
  month = {8},
  publisher = {American Physical Society},
  doi = {10.1103/PhysRevD.110.030001},
  url = {https://link.aps.org/doi/10.1103/PhysRevD.110.030001}
}

@article{Artoisenet:2012st,
    author = "Artoisenet, Pierre and Frederix, Rikkert and Mattelaer, Olivier and Rietkerk, Robbert",
    title = "{Automatic spin-entangled decays of heavy resonances in Monte Carlo simulations}",
    eprint = "1212.3460",
    archivePrefix = "arXiv",
    primaryClass = "hep-ph",
    reportNumber = "NIKHEF-2012-021, CERN-PH-TH-2012-329",
    doi = "10.1007/JHEP03(2013)015",
    journal = "JHEP",
    volume = "03",
    pages = "015",
    year = "2013"
}

@article{Alloul:2013bka,
    author = "Alloul, Adam and Christensen, Neil D. and Degrande, C\'eline and Duhr, Claude and Fuks, Benjamin",
    title = "{FeynRules  2.0 - A complete toolbox for tree-level phenomenology}",
    eprint = "1310.1921",
    archivePrefix = "arXiv",
    primaryClass = "hep-ph",
    reportNumber = "CERN-PH-TH-2013-239, MCNET-13-14, IPPP-13-71, DCPT-13-142, PITT-PACC-1308",
    doi = "10.1016/j.cpc.2014.04.012",
    journal = "Comput. Phys. Commun.",
    volume = "185",
    pages = "2250--2300",
    year = "2014"
}

@article{Bierlich:2022pfr,
    author = "Bierlich, Christian and others",
    title = "{A comprehensive guide to the physics and usage of PYTHIA 8.3}",
    eprint = "2203.11601",
    archivePrefix = "arXiv",
    primaryClass = "hep-ph",
    reportNumber = "LU-TP 22-16, MCNET-22-04, FERMILAB-PUB-22-227-SCD",
    doi = "10.21468/SciPostPhysCodeb.8",
    journal = "SciPost Phys. Codeb.",
    volume = "2022",
    pages = "8",
    year = "2022"
}

@article{Alwall:2014hca,
    author = "Alwall, J. and Frederix, R. and Frixione, S. and Hirschi, V. and Maltoni, F. and Mattelaer, O. and Shao, H. -S. and Stelzer, T. and Torrielli, P. and Zaro, M.",
    title = "{The automated computation of tree-level and next-to-leading order differential cross sections, and their matching to parton shower simulations}",
    eprint = "1405.0301",
    archivePrefix = "arXiv",
    primaryClass = "hep-ph",
    reportNumber = "CERN-PH-TH-2014-064, CP3-14-18, LPN14-066, MCNET-14-09, ZU-TH-14-14",
    doi = "10.1007/JHEP07(2014)079",
    journal = "JHEP",
    volume = "07",
    pages = "079",
    year = "2014"
}

@article{ANUBIS:HNLemu2026,
    author = {Bauer, Martin and Bentham, Rachel and Brandt, Oleg and Erner, Sofie Nordahl and Mullin, Anna and Peng, David and Reymermier, Th\'eo and Swallow, Paul},
    collaboration = {ANUBIS},
    title = "{Projected sensitivity of the ANUBIS detector to heavy neutral leptons}",
    journal = "accepted by Eur.~Phys.~J.~C",
    eprint = {2606.26862},
    archivePrefix = {arXiv},
    primaryClass = {hep-ex},
    year = {2026}
}

@misc{Reymermier:SETANUBISv100,
    author = {Reymermier, Th\'eo and Swallow, Paul},
    title = {{SET-ANUBIS v1.0.0}},
    version = {1.0.0},
    year = {2026},
    publisher = {Zenodo},
    doi = {\href{https://doi.org/10.5281/zenodo.21462101}{10.5281/zenodo.21462101}}
}

@article{erner2026setanubismodularpipelineanubis,
    author = "Erner, Sofie Nordahl and Mullin, Anna and Reymermier, Th{\'e}o and Swallow, Paul",
    collaboration = "ANUBIS",
    title = "{SET-ANUBIS: a modular pipeline for ANUBIS long-lived particle sensitivity studies}",
    eprint = "2607.26112",
    archivePrefix = "arXiv",
    primaryClass = "hep-ph",
    month = "7",
    year = "2026",
    url={https://arxiv.org/abs/2607.26112}
}

@article{ATLAS:2025pak,
    author = "Aad, Georges and others",
    collaboration = "ATLAS",
    title = "{Search for events with one displaced vertex from long-lived neutral particles decaying into hadronic jets in the ATLAS muon spectrometer in pp collisions at s=13{\,}{\,}TeV}",
    eprint = "2503.20445",
    archivePrefix = "arXiv",
    primaryClass = "hep-ex",
    reportNumber = "CERN-EP-2025-062",
    doi = "10.1103/cmql-s9sq",
    journal = "Phys. Rev. D",
    volume = "112",
    number = "9",
    pages = "092001",
    year = "2025"
}

@article{Beacham:2019nyx,
    author = "Beacham, J. and others",
    collaboration = "{PBC}",
    title = "{Physics Beyond Colliders at CERN: Beyond the Standard Model Working Group Report}",
    eprint = "1901.09966",
    archivePrefix = "arXiv",
    primaryClass = "hep-ex",
    reportNumber = "CERN-PBC-REPORT-2018-007",
    doi = "10.1088/1361-6471/ab4cd2",
    journal = "J. Phys. G",
    volume = "47",
    number = "1",
    pages = "010501",
    year = "2020"
}

@article{Bauer:2017ris,
    author = "Bauer, Martin and Neubert, Matthias and Thamm, Andrea",
    title = "{Collider Probes of Axion-Like Particles}",
    eprint = "1708.00443",
    archivePrefix = "arXiv",
    primaryClass = "hep-ph",
    reportNumber = "MITP-17-047",
    doi = "10.1007/JHEP12(2017)044",
    journal = "JHEP",
    volume = "12",
    pages = "044",
    year = "2017"
}

@article{Craig:2018kne,
    author = "Craig, Nathaniel and Hook, Anson and Kasko, Skyler",
    title = "{The Photophobic ALP}",
    eprint = "1805.06538",
    archivePrefix = "arXiv",
    primaryClass = "hep-ph",
    doi = "10.1007/JHEP09(2018)028",
    journal = "JHEP",
    volume = "09",
    pages = "028",
    year = "2018"
}

@article{LHCHiggsCrossSectionWorkingGroup:2016ypw,
    author = "de Florian, D. and others",
    collaboration = "LHC Higgs Cross Section Working Group",
    title = "{Handbook of LHC Higgs Cross Sections: 4. Deciphering the Nature of the Higgs Sector}",
    eprint = "1610.07922",
    archivePrefix = "arXiv",
    primaryClass = "hep-ph",
    reportNumber = "CERN-2017-002-M, CERN-2017-002",
    doi = "10.23731/CYRM-2017-002",
    journal = "CERN Yellow Rep. Monogr.",
    volume = "2",
    pages = "1--869",
    year = "2017"
}

\end{document}